\documentclass[fleqn,usenatbib]{mnras}
\usepackage{newtxtext,newtxmath}

\usepackage[T1]{fontenc}

\DeclareRobustCommand{\VAN}[3]{#2}
\let\VANthebibliography\thebibliography
\def\thebibliography{\DeclareRobustCommand{\VAN}[3]{##3}\VANthebibliography}

\usepackage{lscape}

\newcommand{\XY}[2]{\left[\textrm{#1/#2}\right]}
\newcommand{\FeH}{\XY{Fe}{H}}
\newcommand{\XFe}[1]{\XY{#1}{Fe}}

\newcommand{\kms}{km\,s$^{-1}$}
\newcommand{\Teff}{T_\textrm{eff}}
\newcommand{\logg}{\log g}

\newcommand{\vmic}{v_\textrm{mic}}
\newcommand{\Msol}{\text M_\odot}
\newcommand{\Abund}[1]{A(\text{#1})}

\newcommand{\ch}[1]{\multicolumn 1 c {#1}}

\defcitealias{Mucciarelli2011}{Mu11}	
\defcitealias{Villanova2011}{Vi11}	
\defcitealias{Marino2008}{Ma08}	
\defcitealias{Yong2008b}{Yo08b}	
\defcitealias{Yong2008c}{Yo08c}	
\defcitealias{Ivans1999}{Iv99}
\defcitealias{meszaros_homogeneous_2020}{Me20}

\usepackage{graphicx}	
\usepackage{amsmath}	

\title[Atomic Diffusion and Mixing in M4]{Atomic Diffusion and Mixing in Old Stars VIII: \\ Chemical abundance variations in the globular cluster M4 (NGC\,6121)}

\author[Nordlander et al.]{
T. Nordlander$^{1,2,3}$\thanks{Stromlo Fellow}\thanks{E-mail: Thomas.Nordlander@anu.edu.au}, 
P. Gruyters$^{3,4}$,
O. Richard$^{5}$
and A. J. Korn$^{3}$
\\
$^{1}$Research School of Astronomy and Astrophysics, Australian National University, Canberra, ACT 2611\\
$^{2}$ARC Centre of Excellence for All Sky Astrophysics in 3 Dimensions (ASTRO 3D), Australia\\
$^3$Division of Astronomy and Space Physics, Department of Physics and Astronomy, Uppsala University, Box 516, 75120 Uppsala, Sweden \\
$^4$ Lund Observatory, Box 43, 221 00 Lund, Sweden\\
$^5$ LUPM, Universit\'e de Montpellier, CNRS, CC072, Place E. Bataillon, 34095 Montpellier Cedex, France
}

\date{Accepted XXX. Received 11/12/2023; in original form 29/01/2023}

\pubyear{2023}

\begin{document}
\label{firstpage}
\pagerange{\pageref{firstpage}--\pageref{lastpage}}
\maketitle

\begin{abstract}
Variations in chemical abundances with evolutionary phase have been identified among stars in globular and open clusters with a wide range of metallicities. In the metal-poor clusters, these variations compare well with predictions from 
stellar structure and evolution models considering the internal diffusive motions of atoms and ions, collectively known as atomic diffusion, when moderated by an additional mixing process with a fine-tuned efficiency. 
We present here an investigation of these effects in the Galactic globular cluster NGC\,6121 (M4) ($\FeH = -1.13$) through a detailed chemical abundance analysis of 86 stars using high-resolution ESO/VLT FLAMES spectroscopy. The stars range from the main-sequence turnoff point (TOP) to the red giant branch (RGB) just above the bump. 
We identify C-N-O  and Mg-Al-Si abundance anti-correlations, and confirm the presence of a bimodal population differing by 1\,dex in nitrogen abundance. The composition of the second-generation stars imply pollution from both massive (20--40\,$\rm M_{\odot}$) and asymptotic giant branch stars. 
We find evolutionary variations in chemical abundances between the TOP and RGB, which are robust to uncertainties in stellar parameters and modelling assumptions. The variations are weak, but match predictions well when employing efficient additional mixing. Without correcting for Galactic production of lithium, we derive an initial lithium abundance $2.63 \pm 0.10$, which is marginally lower than the predicted primordial BBN value.
\end{abstract}




\section{Introduction}\label{sec:intro}
The chemical evolution of the Milky Way is believed to be imprinted in the elemental abundance patterns of late-type stars (spectral types F to K). Due to their long lifetimes, these stars are of particular importance when it comes to studying the build-up of elements during the early times of our Galaxy. The chemical composition of the atmospheric layers of such stars is thought to resemble the gas from which they were formed. 
However, observations of globular clusters over the past several decades have revealed a somewhat more complicated picture. Not only have spectroscopic studies revealed a spread in the abundance of light elements in stars of all globular clusters (GCs) \citep[][and references therein]{Gratton2012}, the studies also indicate that there are processes at work in these stars that alter the surface compositions. The sum of these element-separating effects is collectively referred to as atomic diffusion \citep{Michaud1984}. The effects are responsible for an exchange of material between a star's interior and its atmosphere during the main sequence, but can be counteracted as long as convection is efficient. This means that the largest abundance effects are expected for the hotter stars in an old stellar population, i.e. the F-type main-sequence (MS) turnoff-point (TOP) stars. As the stars evolve off the MS, the deepening outer convection zone will restore the original surface composition and null the diffusion effects effectively restoring the stars' original composition.
Prone to proton capture, the element lithium departs from this pciture. As a star ascends the subgiant branch (SGB), the convection zone deepens such that the surface layers dilute with lithium-free material from the interior, causing the surface lithium abundance to drop by an order of magnitude.

A broad search for observational evidence of atomic diffusion (AD) in Population II stars has been driven by the theoretical modelling work of \citet{Michaud1984} and \citet{Richard2002b,Richard2002,Richard2005}. 
The present series of papers (I--VII) gives an overview of the abundance analyses of unevolved and evolved stars in the three globular clusters 
NGC\,6397 \citep{Korn2007,Lind2008,Nordlander2012}, 
NGC\,6752 \citep{Gruyters2013,Gruyters2014}, 
and M30 \citep{gruyters_atomic_2016,gavel_atomic_2021} at metallicities\footnote{We adopt here the customary spectroscopic notations that ${\rm [X/Y]} = \log(N_\text{X} / N_\text{Y}) - \log(N_\text{X} / N_\text{Y})_{\odot}$ for elements X and Y, and that $\Abund X = \log(N_\text{X}/N_\text{H})+12$.} $\FeH  =-2.1$, $\FeH=-1.6$ and  $\FeH=-2.3$, respectively. All three clusters were found to exhibit element-specific evolutionary abundance variations at the level of 0.1--0.3\,dex. 
Beyond this series of papers, studies of lithium in metal-poor stars have provided crucial mass- and metallicity-dependent constraints \citep{gonzalez_hernandez_lithium_2009,melendez_observational_2010,nissen_two_2012,gonzalez_hernandez_6li7li_2019}.
Additional constraints have come from studies of Population I stars, including the solar-metallicity open clusters M67 \citep{onehag_abundances_2014,bertelli_motta_gaia-eso_2018-1,liu_chemical_2019,souto_chemical_2019} and NGC\,2420 \citep{semenova_gaia-eso_2020}, as well as field stars \citep{liu_detailed_2021}, uncovering subtle abundance differences that vary from element to element.

The observed abundance variations are remarkably similar to the ones predicted by stellar evolution models including AD. However, in both cases the observations are not matched by the predictions of AD from first principles, but only when counteracted by an additional mixing mechanism (AddMix). 
This additional transport process beyond the formal extent of the convection zone is also needed to explain the properties of the Spite plateau of lithium \citep{Spite1982} which shows a constant Li abundance in warm Population II stars. Different transport processes have been investigated like those due to mass loss \citep{VaCh95,ViMiRiRi2013}, rotation \citep{DeMa2021} or turbulent mixing \citep{Richard2002,Richard2005}.

The additional mixing is incorporated in the stellar evolution models as an ad-hoc parametric turbulent diffusion coefficient \citep{Richer2000} so that the structure of the model star is modified by mixing a certain depth range. The density dependence ($\rho^{-3}$) is suggested by the Be abundance on the Sun \citep{Proffitt1991}. The only free parameter is the reference temperature $T_0$ that sets the overall efficiency of the AddMix. This family of models uses a shorthand convention T$X$, where $X$ refers to $\log T_0$.
At $T_0$, the AddMix diffusion coefficient $D_T$ is set to 400 times the atomic-diffusion coefficient for helium \citep[see][for the analytic expression of the coefficient]{Richer2000}. With this parameterisation, \citet{Richard2005} were able to reproduce the Spite plateau using a range of models from T6.0 to T6.25. These are also the models explored in this series of papers where we find that the abundance trends in M30 at $\FeH =-2.3$ are best reproduced by the T6.1--6.2 models, NGC\,6397 at $\FeH =-2.1$ by the T6.0 models, and NGC\,6752 at $\FeH =-1.6$ by the T6.2 models. 
In all three cases we find diffusion-corrected Li abundances that are compatible with, but systematically lower than,
predictions of standard Big-Bang Nucleosynthesis (BBN): 
$\Abund{Li} = 2.69 \pm 0.06$ \citep{yeh_impact_2021}. 
These predictions are based on cosmological parameters from observations of the microwave background radiation (CMB) by the \textsc{Planck} satellite \citep{planck_collaboration_planck_2020} and do not involve free parameters fitting predictions to observations of the BBN. 
Given these results, it seems as if the efficiency of AddMix varies with metallicity in a way that is difficult to predict and extrapolate, which could be due to the limited number of observations.
To investigate the validity of this hypothesis we here investigate the Galactic GC M4 (NGC\,6121) at $\FeH =-1.13$ and an age of about 12\,Gyr \citep{bedin_end_2009,vandenberg_ages_2013,jang_multiple_2019}.

M4 is the nearest GC to the Sun, at a distance of 1.8\,kpc \citep{Hendricks2012}. 
Unfortunately, it is located at low Galactic latitude in the Galactic disk behind the Sco-Oph cloud complex, and thus suffers from significant interstellar extinction \citep[$A_V = 1.39$,][]{Hendricks2012} and strong spatial differential reddening \citep{Cudworth1990,Drake1994,Ivans1999}. Deriving effective temperatures ($\Teff$) from photometry thus becomes a cumbersome task. There have nonetheless been several high-quality spectroscopic studies that took these issues into account. \citet{Marino2008,Marino2011} presented evidence for multiple populations along the RGB and HB, while \citet{Monelli2013} demonstrated the presence of two distinct sequences on the RGB. Other high-resolution spectroscopic studies have derived abundances for RGB stars, e.g the lithium content of RGB stars was studied by \citet{Dorazi2010b} and \citet{Monaco2012}. 

M4 has also been the topic of a controversy around the absence or presence of second-generation stars amongst its AGB stars. \citet{maclean_extreme_2016,maclean_agb_2018} analysed high-resolution spectra for a sample of 106 RGB and 15 AGB stars and concluded that there are no second generation stars among the AGB stars. \citet{lardo_multiple_2017}, on the other hand, showed that the AGB is populated by second generation stars using the C$_{U,B,I} = (U-B) - (B-I)$ index, with further support by the spectroscopic and UV photometric study of \citet{marino_spectroscopy_2017}. 

The AD effects on Fe and Li abundances of stars along the evolutionary sequence in M4 was addressed by \citet{Mucciarelli2011}. Although the literature on M4 reveals a tendency that metallicities for TOP stars generally come out lower than those derived for RGB stars (see Table~\ref{Tab:overview}), \citet{Mucciarelli2011} do not find a trend in iron abundance with $\Teff$, i.e. evolutionary stage. They do, however, need AD and very efficient AddMix in order to explain the observed evolution of Li in the cluster. Given the small size of the expected AD trends for Fe and the temperature-sensitivity of Fe, it is not unlikely that the AD effect on Fe would go undetected. As AD affects all elements, we here revisit M4 and derive abundances for 14 elements, to determine whether AD signatures are present.

This paper is organised as follows: The observations are shortly summarised in Sect.~\ref{sec:obs}. In Sect.~\ref{sec:analysis} we outline the derivation of the stellar parameters and discuss our methodology. The results are presented in Sect.~\ref{sec:results} followed by a scientific discussion in Sect.~\ref{sec:Discussion}. We conclude the paper with a summary in Sect.~\ref{sec:summary}.

\begin{table}
\caption{Spectroscopic metallicity determinations of evolved and unevolved stars in M4.}
\label{Tab:overview}
\centering
\begin{tabular}{lccc}
\hline
Reference & $\FeH$ & \# stars & Evolutionary states \\
\hline 
\citet{Mucciarelli2011} & $-1.08$ & 35 & TOP \\
\citet{Mucciarelli2011} & $-1.12$ & 52 & SGB/RGB \\
\citet{Monaco2012} & $-1.31$ & 71 & MS \\
\citet{Monaco2012} & $-1.17$ & 10 & SGB/RGB \\
\citet{Malavolta2014} & $-1.16$ & 1869 & MS/SGB \\
\citet{Malavolta2014} & $-1.07$ & 332 & SGB/RGB \\
\citet{spite_abundances_2016} & $-1.20$ & 71 & TOP \\
\citet{spite_abundances_2016} & $-1.10$ & 10 & SGB \\
\citet{Marino2008} & $-1.07$ & 105 & RGB \\
\citet{marino_spectroscopy_2017} & $-1.20$ & 17 & AGB \\
\citet{wang_sodium_2017} & $-1.14$ & 68 & RGB \\
\citet{wang_sodium_2017} & $-1.18$ & 19 & AGB \\
\citet{maclean_agb_2018} & $-1.17$ & 106 & RGB \\
\citet{maclean_agb_2018} & $-1.20$ & 15  & AGB/HB \\
This work & $-1.20$ & 34 & TOP \\
This work & $-1.13$ & 35 & RGB \\ 
\hline
\end{tabular}
\end{table}



\section{Observations}\label{sec:obs}
We used high-resolution spectroscopic observations of 86 stars in M4 that were obtained with FLAMES/GIRAFFE \citep{Pasquini2003} under ESO programme 081.D-0356, i.e.\ from the same data set as used in \citet[hereafter \citetalias{Mucciarelli2011}]{Mucciarelli2011}. 
The spectroscopic targets were selected by \citet{Lovisi2010}, with a selection informed by proper motions and radial velocities. They used proper motions from \citet{anderson_ground-based_2006}, which clearly separated cluster members from field stars; membership was identified by selecting stars with differential proper motions within 8.5\,mas\,yr$^{-1}$ of the cluster average. \citet{Lovisi2010} determined a cluster mean radial velocity based on the giant stars of $71.25 \pm 0.43$\,\kms ($\sigma = 4.08$\,\kms), and all targets used in this work were found to lie within the envelope of the Gaussian fit to the histogram.
In addition, we have used astrometry from Gaia DR3 \citep{gaia_collaboration_gaia_2021,lindegren_gaia_2021} to reject unresolved binaries and non-members. We identified an additional six stars (IDs 42574, 50403, 53956, 46510, 30922 and 41899) with a poor astrometric solution ($\text{RUWE} > 1.4$) indicating potential binarity, discrepant parallax ($|\varpi - \bar{\varpi}| > 5 \sigma_\varpi$), or having a second resolved source within $1\arcsec$.

\begin{figure}
\begin{center}
\includegraphics{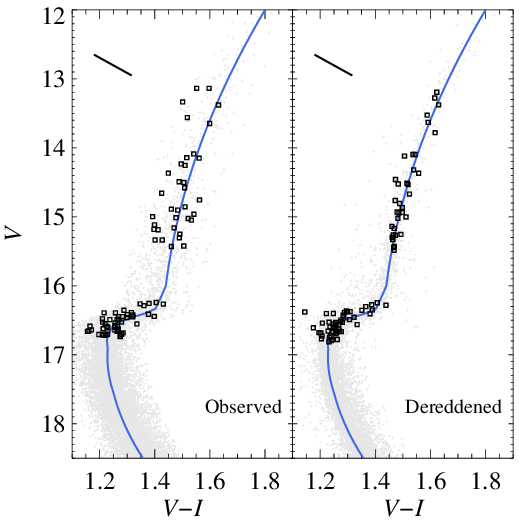}
\caption{Observed $(V-I)$--$V$ colour-magnitude diagram of M4 before and after correcting for differential reddening. The spectroscopic targets are marked by black squares. The reddening vector is given by the solid line in the top left corner. The solid blue line indicates the fiducial sequence. }\label{Fig:CMD}
\end{center}
\end{figure}

The reduced data set used in this work were obtained from the Paris-GIRAFFE archive \citep{Royer2012}. 
Observations used the HR15N, HR18 and HR22 settings, providing the \ion{Li}i doublet at 6707.8\,\AA, several \ion{Fe}{I} lines, the \ion Oi triplet at 7771--7775\,\AA\ and the \ion{C}{I} 9111.8\,\AA\ line, as well as lines for ten other chemical elements we analysed. 
The stars cover a range in evolution from the TOP to the RGB and are shown in the $(V-I)$--$V$ colour magnitude diagram (CMD) in Fig.~\ref{Fig:CMD}. Reduced broadband $UBVI$ photometry of M4 was kindly provided by Y.\ Momany (priv. comm.; see \citealt{Momany2003} for further information; the same photometry was previously used by \citealt{Marino2008}).

\section{Analysis}\label{sec:analysis}
\subsection{Stellar evolutionary models} \label{subsec:isochrones}
We computed a grid of stellar evolutionary models to construct isochrones, using the Montr\'eal-Montpellier stellar evolution code \citep{Richard2002}. The physics of these models are identical to those used previously in this series of papers \citep[see][and discussion in the Introduction]{Richard2005,Korn2007}.

We used the \citet{GrNo93} solar composition and the \citet{TuRiMiIgRo98} solar calibration.
We adopted an initial chemical composition with $Y=0.2382$, $\FeH =-1.1$ and $\rm [\alpha/Fe] = +0.3$ (corresponding to $Z = 0.00268$).
We interpolated the isochrones to an age of 12\,Gyr, consistent with estimates from the white dwarf cooling sequence and isochrone fits to the main-sequence turnoff and horizontal branch \citep{bedin_end_2009,vandenberg_ages_2013,jang_multiple_2019}. 
We calculated models with AD and AddMix efficiencies of $\log T_0 = 6.0$, 6.2, 6.25 and 6.3, with masses in the range 0.55--0.87$\,\Msol$, covering evolutionary states from the lower main sequence to the base of the RGB. We also calculated models with no atomic diffusion, covering masses in the range 0.55--0.874$\,\Msol$, and note that the absence of diffusion of He into the stellar core requires a higher stellar mass by about 0.01\,$\Msol$ to reach the TOP at the same age.

\subsection{Photometry}\label{subsec:phot}

The line of sight towards M4 is heavily affected by interstellar dust due to its location behind the Sco-Oph cloud complex. 
This results in significant differential reddening across the face of the cluster \citep{Ivans1999}, at an average level of $A_V = 1.39$ or $E(B-V) = 0.37$ \citep{Hendricks2012}. 
Estimates of the peak-to-peak differences within a distance of 10' of the cluster centre ranges from $\delta E(B-V) \geqslant 0.05$ \citep{Cudworth1990} to $\delta E(B-V) = 0.25$ \citepalias{Mucciarelli2011}. 
The effect of differential reddening is an apparently broader evolutionary sequence in the cluster CMD than expected from photometric uncertainties alone. This broadening will depend on the angle between the sequence and the reddening vector, e.g., $A_V/E(V-I)$, which is illustrated in Fig.~\ref{Fig:CMD} together with the observed and dereddened cluster CMD. 

To correct for the spatially differential reddening across the face of the cluster, we followed a method similar to that of \citet{milone_acs_2012} and \citet{donati_ngc_2014} and previously applied to M4 by \citet{lardo_multiple_2017}. 
We determined a fiducial sequence by eye, and derived the selective extinction in $(V-I)$ in $\Teff$-$V$ space by comparison to a 12\,Gyr isochrone (described in Sect.~\ref{subsec:isochrones}). 
We derived $\Teff$ values from $V-I$ colours using the relations from \citet{Ramirez2005}, which are calibrated on the infrared flux method \citep[IRFM,][]{Blackwell1986}. We derived a mean selective extinction for cluster members of $E(V-I) = 0.63$. 

\begin{figure}
\begin{center}
\includegraphics[width=1\columnwidth]{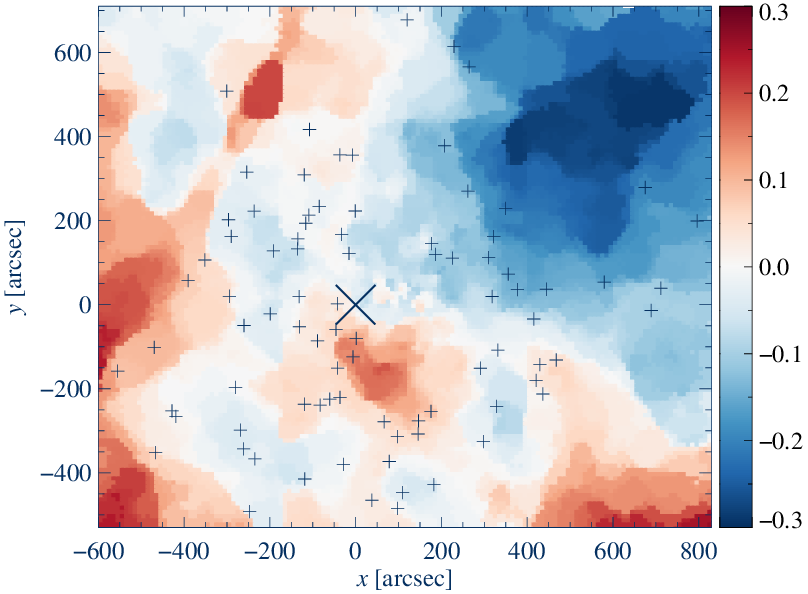}
\caption{Observed $(V-I)$--$V$ reddening map of M4, indicating variations in extinction, $\Delta A_V$, relative to the cluster average. Spectroscopic targets are marked by black crosses, while X marks the centre of the cluster. The blue colour indicates areas with less reddening, while red coloured areas are affected by a higher degree of reddening. The coordinate system is normalised to the cluster centre at R.A. = $16^{\rm h}26^{\rm m}45.12^{\rm s}$, Dec = $26^{\circ}18'35.4''$.
}\label{Fig:dereddened}
\end{center}
\end{figure}

A reference sample of stars was constructed by computing the distance along the reddening vector to the fiducial sequence for all stars located near the main-sequence turnoff point, $16.7 \le V \le 18.3$. For each star, we determine the average reddening by median filtering amongst the nearest $\le35$ neighbouring reference stars within a distance of 60\arcsec\ on the sky. 
The spatial differential extinction in $V$ $(\Delta A_V)$ across the cluster as derived empirically from the comparison to $(V-I)$--$V$ fiducial sequences is visualised in Fig.~\ref{Fig:dereddened}. The surface has been binned to cells of 10\arcsec\ by 10\arcsec. Each cell represents the median $\Delta A_V$ for all stars that fall in the coordinate range and for which a reddening value has been assigned by the method described above. In the map, red and blue colours indicate regions with a reddening value above or below the overall mean reddening of the cluster. The map is qualitatively consistent with the $(B-V)$ reddening maps of \citet{Hendricks2012} and \citet{Monelli2013}.

A corresponding dereddening in the $B-I$ colour index was also derived. As \citet{Ramirez2005} do not provide a $(B-I)$--$\Teff$ transformation, the \citet{Hendricks2012}  transformation factors (see their Table 3) were used to obtain $E(B-I) = 1.06$ from the derived $E(V-I)$. The $B-I$ reddening map is qualitatively similar to that derived from $V-I$. The influence of the differences, between the two maps, on derived stellar parameters will be examined further in Sect.~\ref{sec:photsp}. The good correspondence between the two reddening maps, and the significantly decreased scatter along the RGB in Fig.~\ref{Fig:CMD} validates the accuracy of the dereddening procedure.
We note that while Gaia DR3 provides homogeneous reddening values based on the analysis of BP/RP spectrophotometry \citep{andrae_gaia_2022}, these reddening estimates are marred by significant correlations with the stellar parameters, in particular $\Teff$, and cannot straightforwardly be used.

\subsection{Spectrum synthesis}
We performed an automated spectroscopic analysis using a modified version of the spectrum synthesis code Spectroscopy Made Easy \citep[SME,][]{Valenti1996,Valenti2005,piskunov_spectroscopy_2017}. 
Briefly, stellar parameters ($\Teff$, $\logg$, $\FeH$, and $\vmic$), linelists \citep{Piskunov1995,Kupka1999}, and line and continuum masks are supplied to SME. The code allows non-LTE (NLTE) line formation using precomputed grids of departure coefficients, and uses a grid of MARCS plane-parallel and spherically symmetric model atmospheres \citep{Gustafsson2008}, all with scaled solar abundances, and alpha-enhancement of 0.4\,dex when $\FeH < -1$. SME, like MARCS, adopts the solar chemical composition of \citet{Grevesse2007}. SME performs a numerical comparison between observations and synthetic spectra computed on the fly. The optimum is found through a non-linear optimisation algorithm \citep{Marquardt1963}. 
Rather than a line-by-line approach, we determine all abundances simultaneously by synthesising all three spectral settings for each star. This ensures a consistent abundance table for each star, where all known line blends are explicitly taken into account.

We applied NLTE corrections in our line synthesis routine using precomputed grids of departure coefficients for ten elements: 
lithium \citep{lind_departures_2009}, 
carbon \citep{alexeeva_carbon_2015}, 
oxygen \citep{Sitnova2013}, 
magnesium \citep{osorio_mg_2015,osorio_mg_2016}, 
aluminium \citep{nordlander_non-lte_2017},
silicon \citep{Shi2008}, 
calcium \citep{Mashonkina2007}, 
iron \citep{Bergemann2012,Lind2012} and
barium \citep{Mashonkina1999}. 
We refer the reader to each paper for details on the particular NLTE treatment, and to \citet{piskunov_spectroscopy_2017} for details on the implementation of NLTE departure coefficients in SME. 
While the grids employed in this work are largely proprietary, we note that publicly available grids for several elements have been made available by \citet{amarsi_galah_2020} and \citet{Gerber2023}, and that several online tools are available\footnote{Commonly used online NLTE abundance tools include \url{inspect-stars.com}, \url{nlte.mpia.de}, and \url{spectrum.inasan.ru/nLTE}}.
Other elements and species including CN, \ion K i, \ion{Ti}{ii} and \ion{Ni}i were computed in LTE.

\subsection{Photometric stellar parameters}
\label{sec:photsp}
We used the $V-I$ colour-temperature calibration from \citet{Ramirez2005}, assuming $\FeH = -1.1$ for all stars, and the dereddened $V-I$ photometry as described in Sect.~\ref{subsec:phot}. 
The uncertainty of the dereddened colours can be evaluated by comparing results when dereddening using the independent $V-I$ and $B-I$ reddening maps. Differences are rather small, $7 \pm 34$\,K and $10 \pm 73$\,K for giants and dwarfs, respectively. The scatter about the $V-I$ fiducial sequence corresponds to 45 and 135\,K, for giants and dwarfs. 
The median absolute deviation when executing the nearest neighbour filtering is just 0.032 mag in $V$, corresponding to 0.015 mag in $V-I$. The latter corresponds to a change in $\Teff$ of 36 and 68\,K for giants and dwarfs, respectively. This corresponds well to the scatter in the difference between the two reddening maps, and is comparable to the scatter about the fiducial sequence for giants. Taking into account the statistical uncertainties present, we adopt representative uncertainties of 50 and 100\,K in $\Teff$\ for giants and dwarfs, respectively.

\begin{figure}
\begin{center}
\includegraphics[width=\columnwidth]{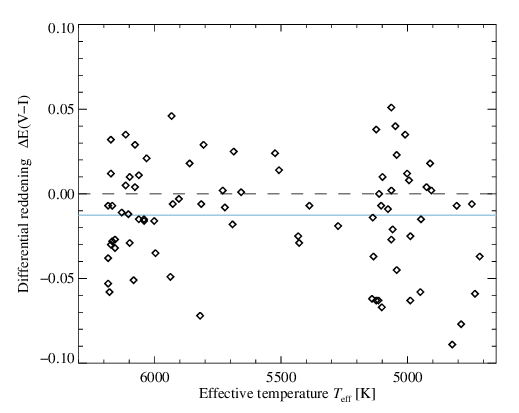}
\caption{Differential reddening $\Delta E(V-I)$ deduced for spectroscopic targets, as a function of their effective temperature. The mean value $-0.013 \pm 0.032$\,mag is indicated by a solid blue line. Differential reddening values are on average slightly more negative for cooler stars (see text for further discussion). }
\label{fig:EVI}
\end{center}
\end{figure}

We show in Fig.~\ref{fig:EVI} that there is little correlation between our inferred stellar parameters and the adopted differential reddening. The average differential reddening is $\Delta E(V-I) = -0.013 \pm 0.032$\,mag, where warm stars with $\Teff > 5800$\,K are slightly more reddened ($-0.010 \pm 0.028$\,mag) than cool stars with $\Teff < 5200$\,K ($-0.018\pm0.037$\,mag). Compared to the mean for the spectroscopic targets, warm stars are thus more highly reddened by just 0.002\,mag and cool stars less by 0.005\,mag. Offsets of $0.005$\,mag in $V-I$ correspond to just 12\,K and 27\,K for warm and cool stars, respectively, and systematic errors in differential reddening of this type are therefore unlikely to be significant.

Compared to the $\Teff$\ scale of \citetalias{Mucciarelli2011}, results are in agreement for turnoff stars (our temperatures are higher by $17 \pm 129$\,K for stars with $\Teff > 5800$\,K), while the values for giants ($\Teff < 5200$\,K) differ by $-58 \pm 21$\,K. Large differences are found on the SGB (intermediate $\Teff$), $-115 \pm 121$\,K.

We derived photometric surface gravities from the dereddened absolute magnitudes, assuming a distance modulus $\mu = 11.28$ \citep{Hendricks2012}, and with bolometric corrections from \citet{Alonso1999,AlonsoEratum}. 
For a 12\,Gyr isochrone with AD (results are not sensitive to the choice of AddMix), we found a mass of 0.84\,$\Msol$ at the TOP and 0.87\,$\Msol$ on the RGB. 
The resulting values of $\logg$ range from 4.2 at $\Teff=6100$\,K (TOP) to 2.2 at $\Teff=4700$\,K (RGB). 
The resulting stellar parameters, along with observational parameters, are given in Table\,\ref{Tab:SPs}.

As outlined in \citet{Gruyters2014}, uncertainties in $\Teff$ are expected to dominate relative errors in $\logg$. For example, an error of $+100$\,K in $\Teff$\ translates into $+0.03$\,dex in $\logg$. An effect of this size on $\logg$\ would require, e.g., an increase in stellar mass by 0.06\,$\Msol$, or an increase in $V$ magnitude by 0.075\,mag. 
We note that our mass estimate is in good agreement with the average of asteroseismic measurements for RGB stars in this cluster,
$0.83 \pm 0.01$ \citep{howell_integrated_2022} or $0.87 \pm 0.01$ \citep{tailo_asteroseismology_2022}. For the 8 and 5 stars that overlap, respectively, the seismic measurements indicate $0.81 \pm 0.09$\, $\Msol$ and $0.79 \pm 0.29$\,$\Msol$. 
Our expected precision of 0.03\,mag in $V$ magnitude and 50 and 100\,K in $\Teff$\ for giants and dwarfs, respectively, thus translates into at most 0.03 and 0.05\,dex in $\logg$ for giants and dwarfs, respectively. A shift in stellar masses, the overall reddening of the cluster, or its distance, would have similar effects on all stars with only minor differential effects.

\begin{table*}
\caption{Dereddened photometry, S/N measured in the continuum in all three spectrograph settings, and stellar parameters (complete table available electronically).}
\label{Tab:SPs}
\centering
\begin{tabular}{lccc ccc ccc}
\hline
ID       & RA          &  Dec        &  $V-I$ & $V$    & S/N &$\Teff$ &$\logg$ & $\vmic$ \\
         & (J2000)     & (J2000)     &  (mag) & (mag)  &     & (K)  &(dex) & (\kms) \\ \hline
M4-8460  &  245.947189 &  -26.374998 &  1.512 & 14.507 & 120 & 4989 & 2.85 & 1.32\\
M4-8777  &  245.937393 &  -26.361475 &  1.462 & 15.257 &  95 & 5115 & 3.21 & 1.24\\
M4-9156  &  245.906998 &  -26.343929 &  1.539 & 14.083 & 115 & 4905 & 2.65 & 1.36\\
M4-13282 &  245.790207 &  -26.391041 &  1.238 & 16.582 &  51 & 6077 & 4.08 & 1.67\\
M4-28007 &  245.805023 &  -26.668879 &  1.504 & 14.949 &  93 & 5058 & 3.04 & 1.28\\
\dots \\
\hline
\end{tabular}
\end{table*}

\subsection{Spectroscopic stellar parameters}
\subsubsection{Microturbulence}
Microturbulent velocities are derived from a set of 17 \ion{Fe}I lines, as the spectra contain only two rather weak \ion{Fe}{ii} lines.
The \ion{Fe}i\ lines span a range in equivalent widths of 20--150\,m\AA\ and 5--80\,m\AA, corresponding 
to $\log W_\lambda / \lambda = -5.5$ to $-4.6$ and $-6.5$ to $-4.9$, for the coolest and hottest stars in our sample. 

We find that dwarfs follow an essentially linear relation of $\vmic$\ increasing with $\Teff$, while giants exhibit $\vmic$\ values decreasing linearly with $\logg$. The RMS scatter about the linear relation for giant stars is just 0.048\,\kms, while that for the dwarfs is considerably higher and amounts to 0.43\,\kms. 
Among the 20 hottest TOP stars ($\Teff > 6080$\,K), which are the faintest and thus exhibit the lowest S/N values, we find a scatter of 0.8\,\kms, and values higher than 3\,\kms\ in several stars. This indicates that we cannot robustly determine $\vmic$\ for all stars in our sample, due to the influence of noise on the weaker lines.

\begin{table*}
\caption{Photometric stellar parameter selection and average photometric stellar parameters for the coadded group-averaged spectra.}
\label{Tab:coadd}
\centering
\begin{tabular}{lrr cc cc }
\hline
Group & \multicolumn 1 c {\# stars} & \multicolumn 1 c {S/N} & $\Teff$-range (K) & log\,$g$-range (dex) & Mean $\Teff$\ (K) & Mean log\,$g$ (dex) \\
\hline
RGB1 & 6  & 208 & 4692--4814 & 2.18--2.43 & $4747\pm47$ & $2.31\pm0.10$ \\
RGB2 & 18 & 308 & 4868--5074 & 2.64--3.08 & $4975\pm60$ & $2.88\pm0.15$ \\
RGB3 & 11 & 267 & 5001--5112 & 3.15--3.28 & $5078\pm36$ & $3.21\pm0.05$ \\
SGB1 & 2  & 81  & 5454--5523 & 3.76--3.81 & $5489\pm48$ & $3.78\pm0.03$ \\
SGB2 & 3  & 86  & 5620--5683 & 3.87--3.92 & $5642\pm35$ & $3.89\pm0.02$ \\
SGB3 & 7  & 122 & 5730--5815 & 3.89--3.98 & $5773\pm37$ & $3.93\pm0.03$ \\
TOP1 & 10 & 142 & 5824--5908 & 3.98--4.08 & $5868\pm27$ & $4.03\pm0.03$ \\
TOP2 & 10 & 129 & 5927--6021 & 3.97--4.14 & $5984\pm34$ & $4.08\pm0.05$ \\
TOP3 & 14 & 137 & 6041--6572 & 4.02--4.20 & $6212\pm152$ & $4.13\pm0.06$ \\
\hline
\end{tabular}
\end{table*}

To alleviate this shortcoming, we also analyse coadded spectra, generated by grouping stars according to their stellar parameters. We opted to perform this grouping according to $\logg$ for the giants but according to $\Teff$ for dwarfs, due to the different slopes in these different parts of the HR diagram.
The characteristics of each group-averaged spectrum is given in Table~\ref{Tab:coadd}.
These $\vmic$\ values follow a similar relation to those determined from individual giant stars, where their slopes as a function of $\logg$ agree to within $1 \sigma$, indicating that the approach is viable. For dwarfs, we find a linear relation as a function of $\Teff$\ with a scatter of just 0.075\,\kms.  Adopting the two relations leads to $\vmic$\ values increasing with $\Teff$ and ranging between 1.2 and 1.7\,\kms\ for the dwarfs ($\Teff > 5500$\,K) while the giants ($\Teff < 5500$\,K, $\logg < 3.8$) have $\vmic$\ values decreasing with $\logg$, between 1.1 and 1.5\,\kms.

\subsubsection{Spectroscopic $\Teff$\ scales} \label{Sec:teffscale}
\begin{figure*}
\begin{center}
\includegraphics[width=0.99\columnwidth]{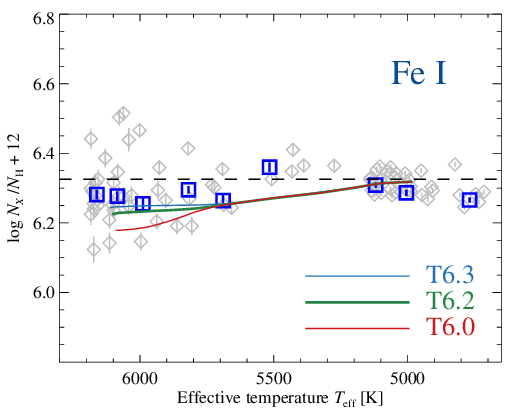}
\includegraphics[width=0.99\columnwidth]{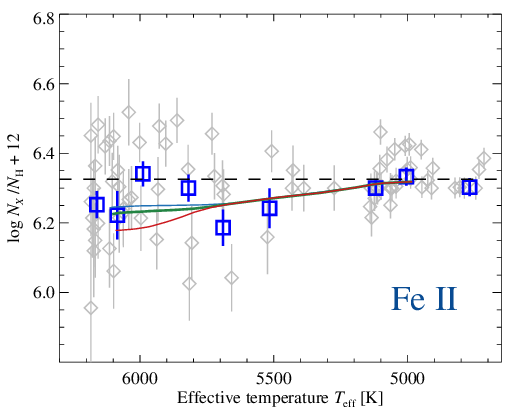}
\caption{Evolutionary abundance trends of iron derived from lines of \ion{Fe}{i} (left) and \ion{Fe}{ii} (right) on the photometric $\Teff$\ scale. The blue squares indicate abundances derived from the coadded group-averaged spectra, while abundances of the individual stars are shown as grey diamonds. Overplotted are predictions from stellar structure models at an age of 12\,Gyr, including atomic diffusion with different efficiencies of additional mixing. 
Note the different behaviour of the two trends, suggesting that the ionisation balance is not fulfilled in the coolest stars.}
\label{Fig:FeI vs Fe II}
\end{center}
\end{figure*}

We compare iron abundances 
derived from lines of \ion{Fe}{i} and \ion{Fe}{ii}, to predictions from evolutionary models with AD in Fig.~\ref{Fig:FeI vs Fe II}. 
Significant differences indicate potential problems with either the photometric stellar parameters or the spectroscopic method. 
In the giants, iron abundances derived from \ion{Fe}{i} lines appear to correlate with effective temperature. Abundances in the coolest stars ($\Teff < 5000$\,K) from \ion{Fe}{i} lines are lower than those from \ion{Fe}{ii} lines by 0.04\,dex. 
The warmer giants, subgiants and dwarfs exhibit better agreement between the ionization stages, with a possible bias of lower abundances of \ion{Fe}{i} than \ion{Fe}{ii} by just 0.03\,dex.

The abundances we derive from lines of \ion{Fe}{ii} are found to be in agreement within the error bars, with predictions from evolutionary models. The abundance trend in titanium as deduced from a \ion{Ti}{II} line is likewise in very good agreement with predictions.
Lines of ionised iron and titanium are known to form under conditions close to LTE, with very small sensitivity to hydrodynamic effects (see Sect.~\ref{Sec:NLTE}), and are rather insensitive to changes in $\Teff$. Our small estimated uncertainties in $\logg$\ and the weak sensitivity to $\vmic$\ likewise indicate these results to be robust.
We thus suspect that the uncovered deviation from ionisation equilibrium is due to inaccuracies in the temperature scale, which the lines of \ion{Fe}{i} are susceptible to. 

The wings of the broad H$\alpha$ line are commonly used in the literature, and are known to yield accurate stellar parameters for dwarf stars \citep{fuhrmann_balmer_1993,barklem_detailed_2002,nissen_sulphur_2007,ruchti_unveiling_2013,amarsi_effective_2018,giribaldi_titans_2021}. 
However, the sensitivity of these lines decreases significantly on the RGB, where \citet{Mucciarelli2011} estimated uncertainties of at least 300\,K. 

Another commonly used method is that of the excitation equilibrium of iron, which should be suitable here as our set of \ion{Fe}{i} lines span a range of excitation energies between 2.5 and 4.5\,eV. 
However, we find that due to the limited number of lines and the limited quality of our spectra among the turnoff stars, differences compared to the photometric temperatures exhibit large scatter with individual differences as large as 550\,K.

We therefore adopt a novel method, and refer to the resulting temperature scale as $T_{\text{eff, ion}}$: we enforce the ionisation equilibrium by matching for each star the iron abundance based on \ion{Fe}{i} lines (in NLTE) to the \emph{average} Fe-trend deduced for all stars from lines of \ion{Fe}{ii}. As this average trend is very similar to that predicted by stellar evolution models including AD with AddMix, on the extreme ends of the temperature scale, where we have the largest number of stars to compare to, we adopt the predicted trend from a T6.2 isochrone at each evolutionary stage.
At low S/N, this fitting method is more robust than the excitation equilibrium.
On average the $T_{\text{eff, ion}}$ scale is cooler than the photometric scale, $T_{\text{eff, phot}}$, by $56 \pm 65$\,K,  $150\pm86$\,K and $29\pm31$\,K for turnoff stars, subgiants, and giants, respectively.

To avoid circular arguments with respect to AD, we also generate a corresponding scale that we refer to as $T_{\text{eff, flat}}$, under the assumption that all stars must indicate the same iron abundance. This $\Teff$\ scale is cooler than $T_{\text{eff, phot}}$, differing by $-9\pm70$\,K, $-87\pm67$\,K and $-34\pm30$\,K for dwarfs, subgiants and giants, respectively. 
Compared to $T_{\text{eff, ion}}$, this temperature scale differs by $+66\pm28$, $+50\pm21$\,K and $-34\pm30$\,K for dwarfs, subgiants and giants, respectively. We will discuss the effect of the different $\Teff$\ scales on the inferred abundances in Sect.~\ref{sec:Tempscales}.

\subsection{Deriving chemical abundances}
We simultaneously determine the abundances of 14 elements, on each of the three $\Teff$\ scales discussed in Sects.~\ref{sec:photsp} and \ref{Sec:teffscale}.
We analyse the light elements lithium using the \ion{Li}i resonance line at 6707.8\,\AA, carbon using the \ion{C}{I} line at 9111.8\,\AA, oxygen using the \ion{O}{I} triplet at 7771-7775\,\AA, magnesium using the \ion{Mg}{I} lines at 7691.5 and 7811.1\,\AA\ and aluminium using the \ion{Al}{I} doublets at 6696-6698\,\AA\ and 7835--7836\,\AA. 
We also determine abundances for the $\alpha$-elements silicon from two \ion{Si}{I} lines, calcium from five \ion{Ca}{I} and two \ion{Ca}{II} lines, and titanium from  the \ion{Ti}{II} line at 6491.5\,\AA. The iron-peak is represented by iron using 18 \ion{Fe}{I} and two \ion{Fe}{II} lines considered separately (where our final abundance analysis is based only on \ion{Fe}{ii}), and nickel using 12 \ion{Ni}{I} lines. Finally, we determine abundances for potassium using the \ion{K}{I} resonance line at 7699\,\AA, and the heavy neutron-capture elements barium using the \ion{Ba}{II} line at 6496.9\,\AA. 

Atomic and molecular line data for full spectrum synthesis were retrieved from the Vienna Atomic Line Database \citep[VALD3,][]{Piskunov1995,Kupka1999,ryabchikova_major_2015}.
For aluminium, we use the TOPbase oscillator strengths and new broadening data from \citet{nordlander_non-lte_2017}. 
We note that newer oscillator strengths are available for some elements, including the \ion O I triplet \citep{Bautista2022} and several lines of \ion{Mg}I \citep{PehlivanRhodin2017}. As this work primarily focuses on differential abundances, the effect of these updated oscillator strengths should be negligible.

As weak CN lines are present over most of the available spectral regions, the nitrogen abundance is also left as a free parameter during the abundance determination for the giant stars. We shall investigate the influence of uncertainties in the abundances of nitrogen on other species in the next section.

\begin{table}
\caption{Abundance sensitivity to stellar parameters. Effects on abundances are shown for the hot and cool ends of our sample, i.e., the coadded spectra RGB1 and TOP3 in Table~\ref{Tab:coadd}.}
\label{Tab:abund sensitivity}
\centering
\begin{tabular}{lrrrrrr}\hline \noalign{\smallskip}
Species &  \multicolumn 2 c {$T_\text{eff}$ $+ 100$\,K}  & \multicolumn 2 c {$\log g$ $+ 0.1$\,dex} & \multicolumn 2 c {$v_\text{mic}$ $+ 0.3$\,\kms} \\ 
    & \ch{TOP}    &   \ch{RGB}    &  \ch{TOP}   &   \ch{RGB}   &   \ch{TOP}   &    \ch{RGB} \\  \hline\noalign{\smallskip}
\ion{Li}{i}  & $  0.077$ & $  0.115$ & $ -0.001$ & $ -0.001$ & $ -0.004$ & $  0.033$ \\
\ion{C}{i}   & $ -0.037$ & $ -0.130$ & $  0.030$ & $  0.051$ & $ -0.013$ & $ -0.021$ \\
N (CN)   & $  \dots$ & $  0.249$ & $  \dots$ & $ -0.021$ & $  \dots$ & $ -0.022$ \\
\ion{O}{i}   & $ -0.057$ & $ -0.092$ & $  0.030$ & $  0.034$ & $ -0.017$ & $ -0.020$ \\
\ion{Mg}{i}  & $  0.039$ & $  0.064$ & $ -0.004$ & $ -0.011$ & $ -0.004$ & $ -0.011$ \\
\ion{Al}{i}  & $  0.051$ & $  0.053$ & $ -0.003$ & $ -0.003$ & $ -0.013$ & $ -0.015$ \\
\ion{Si}{i}  & $  0.034$ & $  0.032$ & $  0.001$ & $  0.005$ & $ -0.002$ & $ -0.026$ \\
\ion{K}{i}   & $  0.072$ & $  0.111$ & $ -0.019$ & $ -0.024$ & $ -0.077$ & $ -0.148$ \\
\ion{Ca}{i \& ii} & $  0.042$ & $  0.073$ & $ -0.002$ & $ -0.009$ & $ -0.048$ & $ -0.099$ \\
\ion{Ti}{ii} & $  0.043$ & $  0.060$ & $  0.036$ & $  0.025$ & $  0.007$ & $ -0.038$ \\
\ion{Fe}{i}  & $  0.073$ & $  0.097$ & $ -0.004$ & $  0.000$ & $ -0.036$ & $ -0.075$ \\
\ion{Fe}{ii} & $  0.015$ & $ -0.029$ & $  0.033$ & $  0.017$ & $ -0.012$ & $ -0.038$ \\
\ion{Ni}{i}  & $  0.061$ & $  0.092$ & $  0.001$ & $  0.008$ & $ -0.011$ & $ -0.070$ \\
\ion{Ba}{ii} & $  0.082$ & $  0.079$ & $  0.018$ & $  0.010$ & $ -0.132$ & $ -0.242$ \\
 \hline
\end{tabular}
\end{table}

\subsection{Abundance uncertainties} \label{sec:uncertainties}
We adopt statistical uncertainties based on $\chi^2$ minimisation, representing photon-noise statistics. These do not take into account uncertainties in continuum placement, nor systematic modelling uncertainties.
We show in Table~\ref{Tab:abund sensitivity} the average effect on the derived abundances from systematic changes to stellar parameters. We note that these are based on a simultaneous abundance analysis of all 14 elements, and therefore include second-order effects due to blends and competition in the molecular equation of state. 

The latter is particularly important as the the presence of weak CN lines in most of the spectral regions produce additional systematic uncertainties, especially in the cooler stars. 
We investigate this uncertainty by executing abundance analyses where we assume different abundances of nitrogen. 
The effect of changing the assumed nitrogen abundance by 0.5\,dex is roughly 0.02\,dex in lithium and 0.01\,dex in magnesium and titanium, while most other elements are unaffected on the level of 0.01\,dex. 
The exception is for carbon and oxygen, which show a somewhat complicated behaviour: The relative abundances of carbon, nitrogen and oxygen determines the relative concentrations of different C-, N- and O-laden molecules in the equation of state. 
In addition, if the abundances of carbon and oxygen are derived while assuming a perturbed nitrogen abundance, then the weak CN lines will present erroneous strengths. In the RGB stars, the CN lines tend to dominate the $\chi^2$ minimization as a function of carbon abundance, with much larger influence than the atomic carbon lines themselves. Setting the nitrogen abundance as a free parameter circumvents this problem, and allows an accurate fit of the atomic carbon lines independent of the CN line strengths. Additionally, the adopted nitrogen abundance does not seem to control the concentration of free atomic carbon and oxygen, especially in the atmospheric layers where the observed  high-excitation lines form, as the influence on the strength of the atomic lines of carbon and oxygen is similar to the effects on lines of magnesium and titanium.
We thus conclude that potential uncertainties in the abundance of nitrogen do not significantly affect results for other species.

\subsection{NLTE and hydrodynamical effects} \label{Sec:NLTE}

\subsubsection{NLTE effects}\label{sec:NLTE}
\begin{figure}
\begin{center}
\includegraphics[width=\columnwidth]{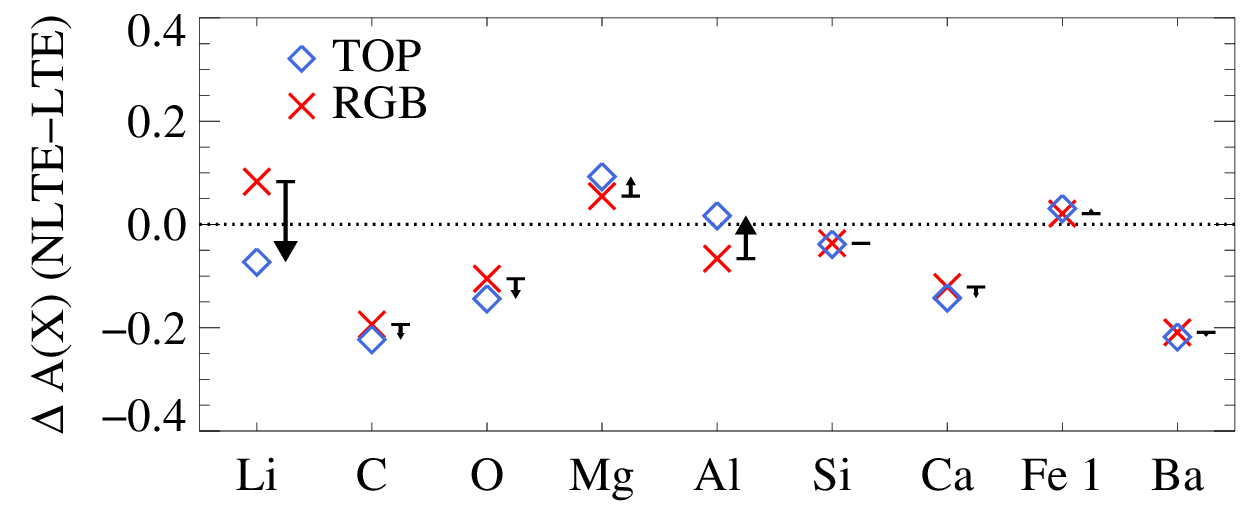}
\caption{NLTE effects, taken as the average difference between NLTE and LTE abundance analyses of TOP (blue diamonds) and RGB stars (red crosses). Arrows indicate the net effect on abundance differences $\Delta \Abund X (\text{TOP}-\text{RGB})$, with positive values represented by upward arrows. 
}
\label{Fig:NLTE}
\end{center}
\end{figure}

We performed our abundance analysis using departure coefficients directly implemented in the spectrum synthesis code.
In order to estimate the magnitude of the NLTE effects, we also derived abundances using LTE line synthesis.
The differences in derived abundances using NLTE line synthesis compared to the LTE case, $\Delta \Abund X (\text{NLTE} - \text{LTE})$, for TOP / RGB stars are illustrated in Fig.~\ref{Fig:NLTE}. The values are typically 
$-0.07$ / $+0.08$\,dex (lithium), 
$-0.22$ / $-0.19$\,dex (carbon), 
$-0.14$ / $-0.10$\,dex (oxygen), 
$+0.09$ / $+0.05$\,dex (magnesium), 
$+0.02$ / $-0.07$\,dex (aluminium)
$-0.04$ / $-0.04$\,dex (silicon), 
$-0.14$ / $-0.12$\,dex (calcium), 
$+0.03$ / $+0.02$\,dex (neutral iron),  and
$-0.22$ / $-0.21$\,dex (barium). 
Save for lithium and aluminium, effects are similar on dwarfs and giants. 

Titanium abundances were derived from a single \ion{Ti}{II} line likely formed under near-LTE conditions, hence we assume the abundances are unaffected by NLTE \citep{Bergemann2011}.
We have examined NLTE corrections for potassium by interpolating the grid of NLTE abundance corrections by \citet{Takeda2002}. We find corrections of $-0.67$\,dex and $-0.57$\,dex for stellar parameters representative of our TOP and RGB stars. 
For nickel, we note the recent publicly available grid by \citet{Gerber2023}, which appeared after the initial submission of this work. An earlier version of this NLTE model was presented by by \citet{bergemann_solar_2021}, and applied to the solar nickel abundance by \citet{magg_observational_2022}. They found that NLTE effects for diagnostic lines were of the order 0.01\,dex, but noted that departures from LTE may be underestimated due to the lack of accurate photoionisation and collisional data \citep{bergemann_solar_2021}. These have since been updated in the work of \citet{Gerber2023}.

\subsubsection{3D hydrodynamic effects}\label{sec:3D}
Three-dimensional (3D) hydrodynamic model atmospheres differ from their 1D counterparts in typically having steeper average temperature stratification, as well as exhibiting horizontal inhomogeneities (granulation). The former will tend to exaggerate differences in line formation under LTE, as this is dictated by the Planck function. In NLTE, however, line formation depends primarily on the average radiation field, which is much more similar in the 1D and 3D cases. This point is illustrated in, e.g., \citet[see their Fig.~6]{Bergemann2012}. We therefore investigate differences between both 3D LTE and 3D NLTE, and 1D LTE modelling, where this is available in the literature.

Detailed comparisons of LTE line formation in 1D and 3D have been performed by \citet{DobrovolskasPhD} and \citet{Dobrovolskas2013}. They compared synthetic spectra of models whose stellar parameters broadly agree with our TOP and RGB groups, and find generally vanishing 3D corrections for weak synthetic lines similar in excitation potential (but not necessarily strength) to those analysed in this work. 

None of the relevant elements exhibit 3D corrections larger than 0.05\,dex, and differential effects, $\Delta \Abund X (\text{TOP} - \text{RGB})$, are found to be less than 0.05\,dex. 
Their tabulations (V.\ Dobrovolskas, priv comm.) indicate that highly excited lines of neutral carbon and oxygen have 3D--1D abundance corrections of 0.00\,dex at the TOP, and small ($-0.03$\,dex) corrections on the RGB. We note however that some of the lines of both elements used in this work are on the verge of saturation (equivalent widths, $EW$, up to 60\,m\AA). Additionally, the excitation potential of the oxygen triplet is higher than what they investigated, which may result in a larger negative correction for weak lines in RGB models.
The neutral lines of silicon, again on the verge of saturation ($EW \sim 50$\,m\AA), exhibit identical $+0.05$\,dex 3D corrections for the TOP and RGB models.
Magnesium and aluminium ($EW \sim 50$\,m\AA) were examined only for an RGB model, indicating slight 3D corrections ($+0.04$ and $+0.03$\,dex). 
The ionised lines of titanium ($EW < 50$\,m\AA) and iron ($EW < 40$\,m\AA) exhibit slight positive 3D corrections in the TOP model ($+0.03$\,dex), but essentially none in the RGB model. The neutral lines of iron and nickel, as well as the neutral and ionised lines of calcium and the ionised line of barium analysed in this work come in a wide range of strengths, most of which are unsuitably strong for these estimates.
Finally, the very weak lines of the CN molecule exhibited essentially no 3D corrections.

These 3D corrections should be seen as indicative of the related uncertainties, rather than quantitative, for two reasons. Firstly, lines analysed in this work are often saturated rather than weak, which affects the contribution functions.

Full 3D NLTE calculations have been performed in the analysis of very metal-poor TOP stars by \citet{Lind2013} for lithium, sodium and calcium. 
Their NLTE corrections for calcium may be compared to the 1D case presented by \citet{Mashonkina2007}, whose corrections are adopted in this work. While these two studies use different atomic models for calcium, they adopt the same inelastic hydrogen collision rates based on \citet{Drawin1968}, rescaled using $S_\text{H} = 0.1$. 
In both studies, the excitation and ionisation equilibria are fulfilled, indicating that NLTE-corrected abundance analyses perform similarly well under 1D and 3D.
The four lines in common for the two studies have NLTE corrections which are either comparable, or significantly larger in the 3D case, indicating that 3D LTE line formation may in fact be less physically realistic than 1D NLTE.
%
%
%

Full 3D NLTE calculations have also been performed for an ultra-metal poor RGB star by \citet{nordlander_3d_2017} for lithium, sodium, magnesium, aluminium, calcium and iron.
They find similar abundances under 1D NLTE and 3D NLTE, with typically stronger NLTE effects in 3D. 
For these six elements they found that either the NLTE effects are mostly negligible (lithium and sodium) with very similar results in 3D NLTE and 1D NLTE or they are substantial (magnesium, aluminium, calcium and iron) with significantly larger abundance corrections in 3D NLTE by as much as 0.3\,dex.
Again, their results indicate that when NLTE effects are expected to be large, they are likely to dominate over the 3D effects such that 
1D NLTE synthesis is preferable over 3D LTE synthesis.

A comprehensive grid of 3D NLTE abundance corrections has been calculated for lithium by \citet{wang_3d_2021}. We have derived indicative abundance corrections through using a representative equivalent width, and calculated (3D NLTE)--(1D LTE) abundance corrections using their provided routines. 
The abundance corrections are rather constant for our stars and range between $-0.07$ and $-0.05$ for the entire sample. While this agrees very well with our 1D NLTE abundance corrections for the warmer stars, these corrections are of the opposite sign to 1D NLTE corrections for the cooler stars, implying that our RGB abundances of lithium are overestimated by as much as 0.15\,dex. 

Grids of 3D NLTE abundance corrections exist for atomic lines of carbon and oxygen \citep{amarsi_carbon_2019-1}. 
For carbon, 3D NLTE abundance corrections range from $-0.05$ for our TOP stars, to $-0.09$ for our RGB stars; this implies abundances are underestimated for dwarfs by roughly 0.17\,dex and for giants by 0.10\,dex. 
For oxygen, corrections are $-0.16$\,dex for both TOP and RGB stars, in very close agreement to our 1D NLTE calculations for dwarfs but lower by 0.06\,dex for giants.



\section{Results}\label{sec:results}
In what follows, the derived chemical abundances are addressed for the sample of 86 stars. The focus will predominately lie on the abundances derived from the group-averaged spectra as we consider the S/N for the warmer TOP stars too low to derive accurate per-star abundances from weak lines, as can be seen in the large scatter we find for the TOP stars compared to the RGB stars.
Results are based on our preferred photometric temperature scale, $T_{\text{eff, phot}}$. 
The stellar parameters and abundance results for group-averaged spectra are presented in Table~\ref{Tab:coadd-results}.

\subsection{Abundance variations}
Abundances derived from the group-averaged spectra generally appear to increase gradually toward lower $\Teff$ values.
The average trends are defined as 
$\Delta \Abund X = \overline{X}_{\text{TOP}} - \overline{X}_{\text{RGB}}$, where $X$ is the investigated element, $\overline{X}_{\text{TOP}}$ the average abundance of the three TOP groups, and $\overline{X}_{\text{RGB}}$ the average abundance of the three RGB groups. 
The difference in stellar parameters between the two groups is about 1000\,K in $\Teff$\ and 1.3\,dex in $\log g$. 
The significance of the trend is based on the standard deviation in the two groups. 
The abundance trends are of the order 0.1\,dex, although some elements such as carbon, oxygen, aluminium and barium seem to exhibit stronger trends ($>0.2$\,dex). 
Although the individual trends are of low significance (1--2$\sigma$), the fact that we find consistent trends in different elements is intriguing. We will address the interpretation of these trends in Sect.~\ref{sec:Discussion}. For now we will continue by describing the various elemental abundances.

\subsection{Lithium} \label{sec:lithium}
The line doublet at 6707.8\,\AA\ used to derived the Li abundance consist of two fine-structure components, separated by merely 0.15\,\AA\ and thus unresolved at the resolution of GIRAFFE $(R=17\,000)$. Our atomic data takes both fine structure (and isotopic splitting) into account. 

The abundances are primarily sensitive to $\Teff$\ since Li is mostly ionised in these stars. Given our $\Teff$\ precision of 100 and 50\,K in dwarfs and giants, we estimate corresponding systematic abundance uncertainties of 0.08 and 0.06\,dex. This dominates the systematic error budget over those due to uncertainties in gravity, metallicity and microturbulence which are only of order 0.01\,dex.

We find the highest lithium abundances in the TOP stars, $\Teff > 5900$\,K, which can be identified with the field star Spite plateau. Their mean abundance $A(\text{Li}) = 2.32 \pm 0.10$ is perfectly consistent with that indicated by the coadded group-averaged spectra, $2.32 \pm 0.03$ as well as with the study by \citetalias{Mucciarelli2011}, who derived $A(\text{Li}) = 2.30 \pm 0.02$ ($\sigma = 0.10$). We will discuss the evolution of Li in greater depth in Sect.~\ref{sec:evol-lithium}.

\subsection[alpha and iron-peak elements]{$\alpha$ and iron-peak elements} \label{sec:aip}
Abundances of Si, Ca, Ti, Fe and Ni all show weak abundance trends based on the group-averaged spectra. The sizes of the trends for Ca, Ti, Fe and Ni are summarised in Table~\ref{Tab:coadd-trends}. 
The most significant trends are found for calcium and nickel (but see discussion below), both of which are well behaved and significant on the 2$\sigma$ level.
The influence of errors in stellar parameters (see Table~\ref{Tab:abund sensitivity}), given our estimated uncertainties, indicates that these systematic uncertainties cannot have spuriously created the trends. 

The trends for silicon, titanium and iron are somewhat less compelling. The latter are deduced from one weak \ion{Ti}{II} and two weak \ion{Fe}{II} lines, meaning that they may be susceptible to the limited data quality of the spectra. As the S/N degrades towards the warmer end of the temperature range, the abundance scatter increases in line with it. 
This leads to a less precise TOP average abundance and at most marginally significant trends: $\Delta(\text{TOP}-\text{RGB}) = -0.06 \pm 0.06$, $ -0.07\pm0.07$ and $ -0.11\pm0.11$\,dex for silicon, titanium and iron, respectively. The abundance trend in iron is somewhat smaller than those between MS and RGB stars found in the literature (see Table~\ref{Tab:overview}). 
However, our abundance trend in iron differs from the null result reported by \citetalias{Mucciarelli2011} who analysed the same spectra. Differences in stellar parameters ($17\pm129$\,K for dwarfs, $-58\pm21$\,K for giant stars) are not sufficient to explain the difference, which is more likely to stem from our use of group-averaged spectra and lines of ionised species, while \citetalias{Mucciarelli2011} analysed lines of neutral species (in LTE) of individual stars.
We note that since the abundances of Fe and Ti are derived from lines of the ionised (majority) species, the trends are rather robust to errors in the $\Teff$ scale, as well as essentially immune to 3D and NLTE effects (see Sect.~\ref{sec:3D} and \ref{sec:NLTE}). The same holds for Si. 

\begin{table*}
\caption{Average abundances based on the coadded spectra and obtained at two effective temperature points.}
\label{Tab:coadd-trends}
\centering
\begin{tabular}{lcccrrrrr}\hline
Group & $T_\text{eff}$ & $\log g$ & $\vmic$ & \ch{$\Abund{Mg}$} & \ch{$\Abund{Ca}$}  & \ch{$\Abund{Ti}$}  & \ch{$\Abund{Fe}$}  & \ch{$\Abund{Ni}$} \\ 
	      & (K)     & (cgs) & (km\,s$^{-1}$) & \ch{NLTE} & \ch{NLTE} & \ch{LTE} & \ch{NLTE} & \ch{LTE} \\ \hline
TOP       & 6005 & 4.08 & 1.61 & $6.70\pm0.01$ & $5.38\pm0.03$ & $3.98\pm0.10$ & $6.25\pm0.07$ & $4.99\pm0.04$ \\
RGB       & 4927 & 2.80 & 1.33 & $6.75\pm0.01$ & $5.45\pm0.01$ & $4.09\pm0.04$ & $6.32\pm0.02$ & $5.07\pm0.01$\\
\hline
$\Delta(\text{TOP}-\text{RGB})$  & 1078 & 1.28 & 0.28 & $-0.05\pm0.01$ & $-0.07\pm0.04$ & $-0.11\pm0.11$ & $-0.07\pm0.07$ & $-0.08\pm0.04$\\
 \hline
\end{tabular}
\end{table*}

\subsection{Light elements}
\subsubsection{C, N, and O}
\label{Sec:CNO}
\begin{figure*}
\centerline{
	\includegraphics[]{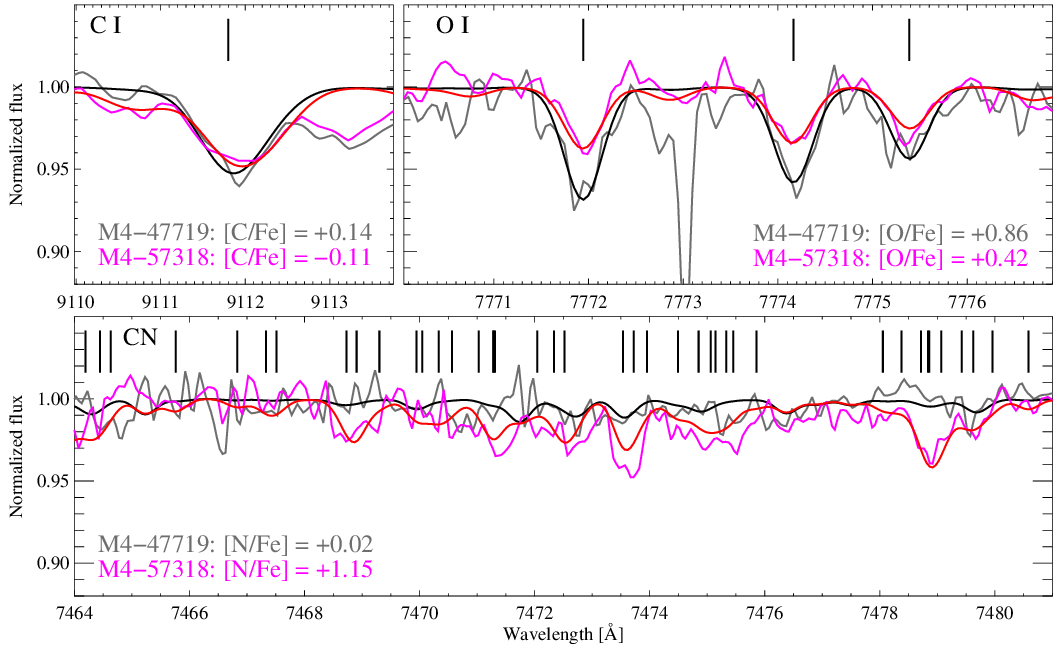}
}
\caption{Comparison of the spectra of two RGB stars with similar stellar parameters, but very different abundances of carbon, nitrogen and oxygen. The synthetic (observed) spectra of M4-47719 are shown in black (grey) and for M4-57318 in red (magenta). Features of \ion C i, \ion O i and the CN molecule are indicated by vertical bars in each panel. The strong spurious feature at 7773\,\AA\ in M4-47719 was automatically flagged and ignored in the analysis of the spectrum.}
\label{fig:CNOspectra}
\end{figure*}

We derived carbon and oxygen abundances simultaneously from neutral atomic lines, along with nitrogen abundances from a large number of vanishingly weak CN features in the cool giants (see the discussion in Sect.~\ref{sec:uncertainties}).
Although the individual CN lines are weak, their combined influence on the $\chi^2$ minimisation appears sufficient to broadly classify stars as either rich or poor in nitrogen. We find the best constraints on the N abundance in our coolest RGB stars. 
We do not fit nitrogen abundances in the dwarfs, given the weakness of the CN features. 
We illustrate the variation found in these features in Fig.~\ref{fig:CNOspectra}, comparing the spectra of stars with similar stellar parameters but different abundances of carbon, nitrogen and oxygen.

Amongst the giant stars, carbon and oxygen abundances both exhibit a tip-to-tip scatter of 0.5\,dex ($\sigma = 0.12$ and 0.11\,dex). Abundances derived from the group-averaged spectra show similar, strong, trends with evolutionary phase:  $\Delta(\text{TOP}-\text{RGB}) = -0.24\pm0.10$ and $-0.27\pm0.04$ for C and O, respectively. 

Comparing our C abundances, derived in NLTE from the neutral line at 9112\,\AA, with those in the literature derived from CN lines or the CH G-band reveals large offsets. The lowest C abundances were found in rather evolved stars by \citet{Ivans1999}, $\XFe{C} = -0.50$ (or $-0.35$ when adjusted to the solar abundance scale of \citealt{Grevesse2007}), while \citet{Villanova2011} found a slightly higher value, $\XFe{C} = -0.28$. 
Our average value in the RGB stars, $\XFe{C} = 0.05 \pm 0.15$, is significantly higher, and does not appear to change systematically over the intrinsic scatter in the most evolved giants. 
We note that the NLTE correction for the \ion C i line at 9112\,\AA\ in our work typically reduces the abundance by 0.2\,dex. The abundance corrections from \citet{amarsi_carbon_2019-1} are more positive by roughly 0.1\,dex, implying a minor systematic shift in our abundances, thus further increasing the difference with the literature. 
We note however that comparisons of C and N abundances amongst red giants stars crucially depend on the precise evolutionary state of the star, as dredge-up significantly alters these surface abundances in a way that is not well predicted from first principles \citep[e.g.,][]{placco_carbon-enhanced_2014,henkel_phenomenological_2017,lagarde_gaia-eso_2019}. 

The oxygen abundances we find are again somewhat higher than those previously reported in the literature. For our RGB stars, we find a mean $\XFe O = 0.67\pm0.10$\,dex, which is higher than the results of \citet[$\XFe O = 0.56$, or 0.54 on our adopted abundance scale]{Yong2008b}, \citet[$\XFe O = 0.39$]{Marino2008}, \citet[$\XFe O = 0.34$]{Villanova2011} and \citet[$\XFe O = 0.25$]{Ivans1999}. All these authors derived the O abundance from the forbidden [O {\scriptsize I}] line at 6300\,\AA, which in metal-poor stars is understood to be nearly immune to NLTE and 3D effects \citep{amarsi_non-lte_2016-1} and to the blend with Ni that is otherwise influential in solar-metallicity stars \citep{bergemann_solar_2021}. 
We note that on the alternative temperature scale $T_{\text{eff, ion}}$, our average $\XFe O = 0.58\pm0.09$ is in excellent agreement with the results of \citet{Yong2008b}.
This agreement lends support to the accuracy of temperatures derived from this form of the ionisation equilibrium, which thanks to the ease with which the average [Fe/H] can be measured, as compared to abundances of individual lines in the excitation equilibrium, may be suitable in general for spectroscopy of faint cluster members and especially at low S/N and low resolution.

\begin{figure*}
\begin{center}
\includegraphics[]{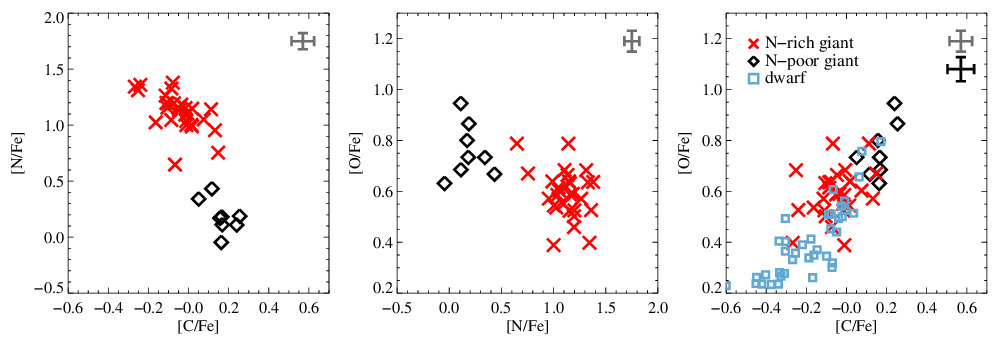}
\caption{Observed correlations between the light elements C, N and O. Red crosses indicate N-rich giants while N-poor giants are shown as black diamonds. We could not determine N-abundances in the dwarfs, and so compare only their abundances of C and O using blue squares. Typical uncertainties on the abundances are shown in the top right corner of the panels, in gray for giants and in black for dwarfs.}\label{Fig:CNO}
\end{center}
\end{figure*}

Abundance ratios are compared in Fig.~\ref{Fig:CNO}. A bimodal distribution of nitrogen abundances among the giants is apparent, with a gap near $\XFe N \sim 0.5$. 
The sample thus splits into two well separated groups having different light-element contents. This behaviour was previously noted by \citet{Marino2008}, and confirmed by \citet{Villanova2011}. Both authors found also that sodium abundances correlated well with nitrogen. From their findings, they concluded that M4 has two distinct chemical populations. We confirm this conclusion and find 10 N-poor giants which belong to the first generation, and 30 N-rich giants belonging to the second generation. 
This indicates that $25 \pm 7$\,\% of our stars belong to the first generation, in good agreement with the study by \citet[20\,\%]{Carretta2013a}, as well as the general consensus that about 1/3 of the stars in a GC belong to the first generation \citep[see e.g.][]{Carretta2010,Gratton2012}. 

Figure~\ref{Fig:CNO} shows the usual anti-correlations between C/O and N, where N-rich stars are characterised by low abundances of C/O, while N-poor stars exhibit high C/O abundances. In the third panel of the figure, the C-O correlation is displayed for giants and dwarfs. The relations are similar to what \citet{Marino2008} and \citet{Villanova2011} have found, which generally supports the N abundances we derive. We calculated the total C+N+O abundance, where possible, and find it to be constant as expected from stellar evolution, with a mean value of $8.32 \pm 0.08$. This is somewhat higher than was found by \citet{Villanova2011} and \citet{Ivans1999}, who derived 8.16 and 8.24 respectively. Table~\ref{Tab:CNO} gives the average abundances for the N-poor and N-rich sub-populations. 
Based on this abundance data, we find that the two sub-populations identified according to $\XFe N$ have significant abundance differences in their abundances of C and O, and possibly Al, 
but no significant differences in Li, C+N+O, Mg, Si, Ca, Ti, Fe, Ni and Ba. 
We will return to this in Sect.~\ref{sec:lights} and \ref{sec:heavies} below.

\begin{table*}
  \caption{Mean abundances of the two M4 sub-populations identified in the giant stars, and their combined mean values compared to literature values.}\label{Tab:CNO}
\begin{tabular}{lrr|c|rrrrrrr}
\hline
Element & \multicolumn 1c{N-rich} & \multicolumn 1c{N-poor} & \multicolumn 1 c {$A(X)_\odot$} & \multicolumn 1 c {M4} &
	\citetalias{Villanova2011} & \citetalias{Mucciarelli2011}$^{a}$  & \citetalias{Marino2008} & 
	Yo08$^{b}$ & \citetalias{Ivans1999} & \citetalias{meszaros_homogeneous_2020}$^{c}$ \\
\hline
$\Abund{Li}$    & $ 1.11\pm0.15$ & $ 1.12\pm0.15$ & $1.05$ & $ 1.11$ & $ 0.97$ & $ 0.92$ &  \dots  &  \dots  &  \dots & \dots \\
$\XFe{C}$       & $-0.04\pm0.11$ & $ 0.17\pm0.07$ & $8.39$ & $ 0.02$ & $-0.28$ &  \dots  &  \dots  &  \dots  & $-0.50$ & $-0.33$\\
$\XFe{N}$       & $ 1.03\pm0.16$ & $-0.05\pm0.34$ & $7.78$ & $ 0.76$ & $ 0.48$ &  \dots  &  \dots  &  \dots  & $ 0.85$ & $0.89$ \\
$\XFe{O}$       & $ 0.59\pm0.12$ & $ 0.76\pm0.11$ & $8.66$ & $ 0.64$ & $ 0.34$ & $ 0.30$ & $ 0.39$ & $ 0.56$ & $ 0.25$ & $0.39$\\
$\Abund{C+N+O}$ & $ 8.31\pm0.07$ & $ 8.36\pm0.09$ & $8.88$ & $ 8.32$ & $ 8.16$ &  \dots  &  \dots  &  \dots  & $ 8.24$ & $8.19$ \\
$\XFe{Mg}$      & $ 0.35\pm0.06$ & $ 0.34\pm0.06$ & $7.53$ & $ 0.35$ & $ 0.47$ &  \dots  & $ 0.50$ & $ 0.57$ & $ 0.44$ & $0.49$ \\
$\XFe{Al}$      & $ 0.51\pm0.06$ & $ 0.36\pm0.14$ & $6.37$ & $ 0.48$ & $ 0.52$ &  \dots  & $ 0.54$ & $ 0.74$ & $ 0.64$ & $0.71$ \\
$\XFe{Si}$      & $ 0.35\pm0.06$ & $ 0.33\pm0.08$ & $7.51$ & $ 0.34$ & $ 0.43$ &  \dots  & $ 0.48$ & $ 0.58$ & $ 0.65$ & $0.46$ \\
$\XFe{K}$       & $ 0.64\pm0.09$ & $ 0.69\pm0.09$ & $5.08$ & $ 0.65$ &  \dots  &  \dots  &  \dots  &  \dots  &  \dots  & $0.11$ \\
$\XFe{Ca}$      & $ 0.27\pm0.04$ & $ 0.27\pm0.08$ & $6.31$ & $ 0.27$ & $ 0.41$ &  \dots  & $ 0.28$ & $ 0.42$ & $ 0.26$ & $0.32$ \\
$\XFe{Ti}$      & $ 0.37\pm0.08$ & $ 0.33\pm0.11$ & $4.90$ & $ 0.36$ & $ 0.33$ &  \dots  & $ 0.32$ & $ 0.41$ & $ 0.30$ & \dots \\
$\FeH$          & $-1.12\pm0.04$ & $-1.13\pm0.07$ & $7.45$ & $-1.13$ & $-1.14$ & $-1.11$ & $-1.07$ & $-1.23$ & $-1.18$ & $-1.02$ \\
$\XFe{Ni}$      & $-0.02\pm0.04$ & $-0.04\pm0.06$ & $6.23$ & $-0.03$ & $-0.01$ &  \dots  & $ 0.02$ & $ 0.12$ & $ 0.05$ & \dots \\
$\XFe{Ba}$      & $ 0.30\pm0.14$ & $ 0.26\pm0.12$ & $2.17$ & $ 0.29$ & $ 0.31$ &  \dots  & $ 0.41$ &  \dots  & $ 0.60$ & \dots \\
\hline
\end{tabular}\\
\textit{(a)}: These $\XFe O$ values refer to abundances from \citet{Lovisi2010}. \textit{(b)}: Data from \citet{Yong2008b} and \citet{Yong2008c}. \textit{(c)}: Data are partly based on \citet{meszaros_homogeneous_2020}. 
\end{table*}

\subsubsection{Mg, Al and K}\label{sec:lights}
Besides the light elements C, N and O, we also derive abundances for magnesium, aluminium and potassium. The Al abundances show large variations of up to 0.5\,dex measured tip-to-tip ($\sigma = 0.11$\,dex). Mg and K show hardly any intrinsic variation, with tip-to-tip differences of just 0.2\,dex ($\sigma = 0.04$ and 0.09\,dex). All elements again show evidence for a gradual increase in abundances with decreasing $\Teff$\ in the group-averaged spectra. The size of the abundance trends of Mg and K is comparable to that found for Si. Al, on the other hand, shows a stronger trend, more comparable to that of carbon or oxygen.

\begin{figure*}
\begin{center}
\includegraphics[]{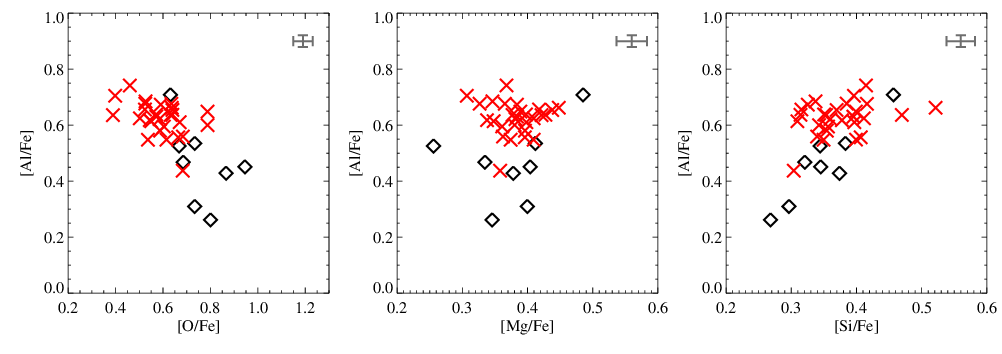}
\caption{Observed correlations between aluminium and the light elements O, Mg and Si in giant stars. The symbols and colours are the same as in Fig.~\ref{Fig:CNO}. The black cross in the top right corner of each panel represents the typical error on the abundances.}\label{Fig:anticor}
\end{center}
\end{figure*}

From studies on chemical populations in globular clusters, we expect magnesium to be anti-correlated with sodium. Unfortunately we do not have information on sodium in our spectra. In massive clusters such as NGC\,6752, NGC\,2808 and NGC 7078, (anti-)correlations linking Mg-Al and Si-Al have also been observed \citep{Yong2005,Carretta2009a}. M4, however, does not show evidence of a Mg-Al anti-correlation in our data. The (anti-)correlations between Al and O, Mg and Si for the giants are illustrated in Fig.~\ref{Fig:anticor} where we make a distinction between N-rich and N-poor stars, defined by a separation at $\XFe N = 0.5$. We find a clear Si-Al correlation and O-Al anti-correlation, both significant on the $3\sigma$ level \citep[taking into account errors on both values, using the IDL routine \textsc{linmix\_err}, see][]{Kelly2007} but no Mg-Al anti-correlation.
%
%

Low Al and Si abundances are only found in the N-poor stars, suggesting that a N-Al correlation is present. 
The average Al abundances for the two sub-populations differ by $0.13 \pm 0.18$\,dex, where the large dispersion in the N-poor group dominates the uncertainty and thus precludes us from drawing any firm conclusion, although an anti-correlation is formally highly significant at $4 \sigma$. 
We find that the stars with the lowest Al abundance are characterised by low abundances of N and correspondingly high C and O in line with several previous studies \citep{Ivans1999, Marino2008, Carretta2013a, nataf_relationship_2019, meszaros_homogeneous_2020}. \citet{Villanova2011}, on the other hand, did not find N-Al or Si-Al correlations and argue that a possible N-Al correlation may be spuriously caused by an unrecognised molecular line (possibly CN) blended with the Al lines at 6696--6698\AA. Our inclusion of two additional \ion{Al}{I} lines at 7835--7836\,\AA\ should help diminish such effects.
By visually inspecting the agreement of the two \ion{Al}i doublets in both N-rich and N-poor stars, we conclude that the variation in Al is real, and that the N-Al and Si-Al correlations are plausible. 
Interesting to note are the large variations in Al and Si observed in the N-poor stars. We will discuss this further in Sect.\,\ref{sec:formation}.

\subsection{Heavy elements}\label{sec:heavies}
We derived abundances for the neutron-capture element barium, whose production in the solar chemical composition is dominated by the s-process \citep{arlandini_neutron_1999,simmerer_rise_2004}. 
We derived NLTE-corrected Ba abundances for all stars, and find a mean value when considering the group-averaged RGB spectra of $\rm [Ba/Fe] = 0.38\pm0.08$, or $0.22\pm0.07$ for the group-averaged TOP stars.
The result for the RGB stars agrees well with three out of four values found in the literature, ${\rm [Ba/Fe]} = 0.60 \pm 0.10$, $0.41\pm0.09$, $0.50\pm0.12$ and $0.32\pm0.04$ from \citet{Ivans1999}, \citet{Marino2008}, \citet{Dorazi2010b} and \citet{Villanova2011}, respectively.



\section{Discussion} \label{sec:Discussion}

\begin{figure*}
\begin{center}
\includegraphics[]{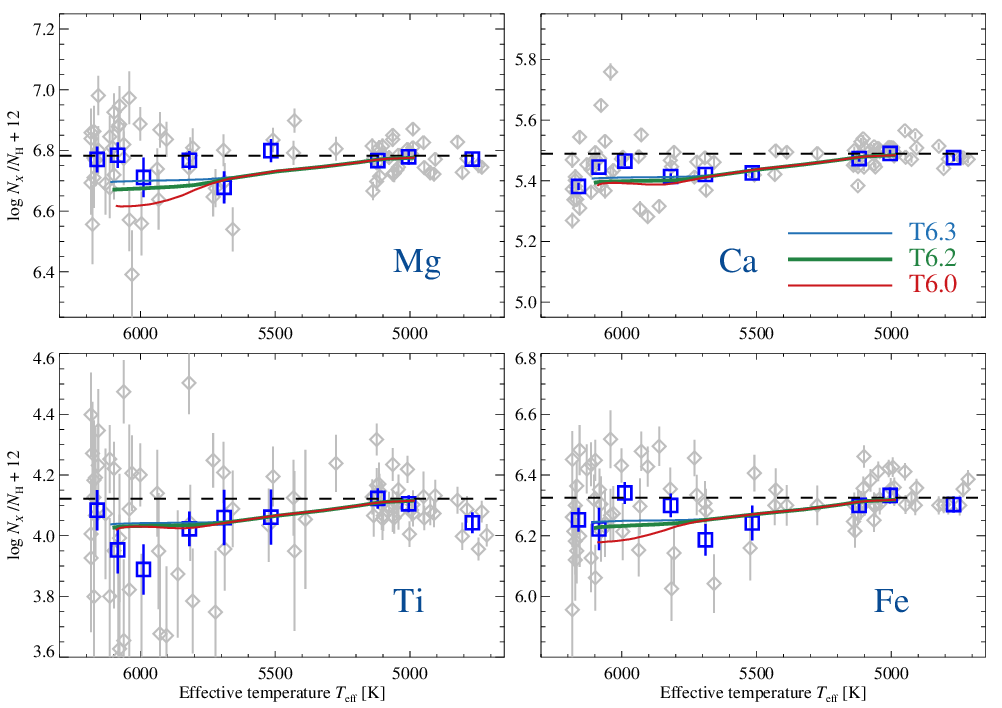}
\caption{Evolutionary abundance trends of Mg, Ca, Ti and Fe. Mg abundances are derived from neutral lines, while Ti and Fe are derived from lines of singly ionised species, and Ca is based on a mixture of lines of both neutral and singly ionised species. The trends are compared to predictions from stellar structure models including AD with additional mixing with different efficiencies, at an age of 12\,Gyr. Horizontal, dashed lines represent the initial abundances of the models, which have been adjusted so that predictions match the observed abundance level of the coolest stars. Blue squares represent results for group-averaged coadded spectra, while gray diamonds represent results for individual stars.}
\label{Fig:AD-trends}
\end{center}
\end{figure*}

\subsection{Atomic diffusion trends}\label{subsec:AD}
Abundance trends with $\Teff$\ for magnesium, calcium, titanium and iron are shown in Fig.~\ref{Fig:AD-trends}. The squares in the figure represent the abundances derived from the coadded group-averaged spectra, while abundances for the individual stars are shown as grey diamonds.
The observed abundances are compared to predictions from stellar evolution models that take into account the effects of atomic diffusion (AD) and additional mixing (AddMix), described in further detail in Sect.~\ref{subsec:isochrones}. 

The abundance trends in Fig.~\ref{Fig:AD-trends} are compared to AD models at three different efficiencies of AddMix. The T6.0 ($\log T_0 = 6.0$) grid of models represents models with low efficiency of AddMix, which has previously been found to well match observations in NGC\,6397 \citep{Korn2007,Lind2008,Nordlander2012}. 
The stronger mixing of the T6.2 models counteracts the effects of AD more efficiently, resulting in weaker abundance trends as compared to the T6.0 models. 
This grid of models with higher efficiency was preferred to explain the trends observed in NGC\,6752 \citep{Gruyters2013,Gruyters2014} and in M30 \citep{gavel_atomic_2021}.
The present results prefer T6.2 over T6.0, primarily on the basis of the shallow abundance variations in Mg and Fe. 
We have also included models with very high efficiency of AddMix (T6.3), but note that these only differ significantly from the T6.2 models in the predicted evolution of lithium. 
The effects on surface abundances in TOP stars differ between the T6.2 and T6.3 models by just 0.02\,dex for elements like carbon, oxygen, magnesium and silicon, but even less for calcium, titanium and iron-peak elements. Effects on lithium however differ by 0.4\,dex, due to the large amount of burning caused by deeper mixing in the T6.3 grid of models. 
We thus prefer the T6.2 models over T6.3 on the grounds of remaining conservative in estimating the effects on lithium (see Sect.\,\ref{sec:evol-lithium}).

Given the consistent appearance of abundance trends in five elements, in good agreement with model predictions regarding both sign and magnitude, it seems unlikely that these trends are a spurious result of errors in measurements and stellar parameters, and modelling shortcomings.
For example, due to the weak temperature sensitivity of lines of singly ionised species, flattening the abundance trend determined in iron would require, e.g., raising temperatures on the TOP by $+450$\,K, or on the RGB by $+250$\,K, in contrast to the estimated uncertainties of 100 and 50\,K, respectively. Similarly, the (formally) required changes on the TOP by $+0.2$\,dex in $\logg$ or $+1.7$\,\kms\ in $\vmic$, or on the RGB by $-0.4$\,dex and $+0.6$\,\kms\ are considered unlikely. Additionally, such changes could not \emph{simultaneously} generate null trends in all five elements.

\subsection{Evolution of lithium} \label{sec:evol-lithium}
\begin{figure}
\begin{center}
\includegraphics[]{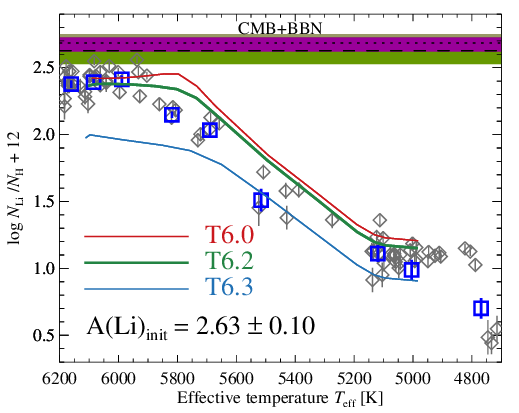}
\caption{Observed lithium abundances, compared to stellar evolution model predictions for different efficiencies of AddMix (see text). Measurements for individual stars are shown as grey diamonds, while squares correspond to the Li abundances derived from coadded spectra. The initial, i.e. diffusion-corrected, abundance of the models, $\Abund{Li} = 2.63 \pm 0.10$, shown by the horizontal dashed line and shaded region, compares well to the predicted primordial lithium abundance, $\Abund{Li} = 2.69$, shown by the dotted horizontal line and shaded region.}\label{Fig:lithium}
\end{center}
\end{figure}

We compare lithium abundances to model predictions in Fig.~\ref{Fig:lithium}. 
While the models of low-efficiency AddMix, T6.0, predict a slight upturn in surface lithium abundances as stars evolve onto the SGB, the higher-efficiency model T6.2 instead predicts a slight decrease. This is a direct consequence of the depths where AddMix operates: In the low-efficiency models, gravitational settling causes the Li abundance to increase with depth below the convective zone during the main sequence, in layers where the temperature is not high enough to destroy Li. When the star evolves off the main sequence, the convection zone expands inwards and the settled material resurfaces \citep[see][]{Korn2006}. 
By contrast, in the higher-efficiency T6.2 models AddMix operates over a larger extent inside the star, without significant deposition of Li below the convection zone (see section 3.4.3 of \citealt{Richard2005} for details). As the convection zone expands inward, lithium-depleted material dilutes the surface composition. In the T6.3 models with highest efficiency of AddMix, lithium is brought directly to regions where temperatures are sufficient for nuclear burning, resulting in strongly depleted surface layers already at the TOP.

This same dilution mechanism is responsible for the rapid decrease in surface lithium abundances along the SGB, during the first dredge-up. 
Firstly, we find that abundances at the TOP-SGB transition, $A(\text{Li}) = 2.19 \pm 0.04$, are lower than those on the TOP, $A(\text{Li}) = 2.40 \pm 0.09$, which disfavours the lithium turn-up predicted by the low-efficiency AddMix model.
Secondly, we find that the observed smooth decrease in lithium abundances during the first dredge-up match predictions reasonably well.
Following the dredge-up, a plateau is reached on the RGB, with an average abundance of $A(\text{Li}) = 1.09 \pm 0.05$. 
Finally, lithium abundances drop sharply on the cool end of the RGB, with three stars exhibiting values significantly below the plateau, averaging $A(\text{Li}) = 0.49 \pm 0.05$. The physics of this extra-mixing episode, likely caused by thermohaline mixing, are not included in our models, but are available and well described elsewhere \citep[e.g.,][]{dearborn_three-dimensional_2006,Charbonnel2010,henkel_phenomenological_2017}.

The evolution of Li is qualitatively consistent with that presented by \citetalias{Mucciarelli2011}. Comparing lithium abundances, their abundances are lower by 0.10\,dex among the TOP stars, on the RGB by 0.17\,dex and after the extra-mixing episode by 0.26\,dex.
The rather large difference for RGB stars cannot be explained by differences in stellar parameters (our $\Teff$\ values are lower by $58 \pm 21$\,K, leading to \emph{lower} abundances by 0.07\,dex) and NLTE corrections (their corrections are more \emph{positive} by roughly 0.10 and 0.15--0.20\,dex at the lower and upper stages on the RGB).

We correct the observed lithium abundances for the predicted amount of depletion using the T6.2 models, resulting in an average (weighted mean) initial lithium abundance of $\Abund{Li}_\text{init} = 2.70 \pm 0.08$ among the TOP stars. This is in fair agreement with the corresponding value recovered from the RGB plateau, $2.59\pm0.07$, with a difference $\Delta \Abund{Li}_\text{init}(\text{TOP}-\text{RGB}) = 0.11$, and for the full sample of stars (excluding the three brightest RGB), $\Abund{Li}_\text{init} = 2.63 \pm 0.10$. We adopt this latter result as our recommended value, and note its close agreement with values determined for NGC\,6752 \citep[$2.58\pm0.10$ or $2.53\pm0.10$,][]{Gruyters2013,Gruyters2014}, NGC\,6397 \citep[$2.57\pm0.10$,][]{Nordlander2012} and M30 \citep[$2.48\pm0.10$,][]{gruyters_atomic_2016}.

Unfortunately, physical shortcomings of the stellar evolution models aside, additional uncertainty stems from the choice of AddMix efficiency. For example, selecting the weaker efficiency T6.0 results in an average $\Abund{Li}_\text{init} = 2.55$ (with $\Delta \Abund{Li}_\text{init}(\text{TOP}-\text{RGB}) = 0.09$ dex) while higher efficiency, T6.25, results in $\Abund{Li}_\text{init} = 2.73$ (with $\Delta \Abund{Li}_\text{init}(\text{TOP}-\text{RGB}) = 0.17$ dex), nevertheless, both values are in agreement with our recommended value. 
Increasing the AddMix efficiency even more to T6.3 results in $\Abund{Li}_\text{init} = 2.97 \pm 0.16$, albeit with considerably larger difference between the initial abundance deduced from TOP and RGB stars, $\Delta \Abund{Li}_\text{init}(\text{TOP}-\text{RGB}) = 0.28$ dex. This is because such high efficiency of AddMix leads to a lithium reduction at the TOP that is dominated by nuclear burning rather than deposition; the lithium gap between the TOP and RGB plateau therefore constrains the maximum efficiency of the mixing to a value less than T6.3.
\cite{Mucciaetal2022} discovered a thin lithium plateau in metal-poor RGB stars, in addition to the lithium plateau. They found that models similar to ours, including AddMix, could reproduce both plateaus with the same value of AddMix throughout evolution, as we obtain for M4 in this work.
We note that the 3D NLTE abundance corrections from \citet{wang_3d_2021} lead to lower abundances derived from the RGB stars by roughly 0.15\,dex, which would significantly worsen the agreement for all models and in particular these higher efficiency models. 

As noted in the previous section, the evolutionary effects of Add\-Mix efficiencies in the range T6.2--T6.3 on elements other than lithium are treacherously indistinguishable. This makes an accurate inference of the initial lithium abundance from RGB stars alone, as proposed by \citetalias{Mucciarelli2011}, difficult as the inferred initial abundances in our case using the T6.2--6.3 models cover the range $\Abund{Li}_\text{init} = 2.63$--2.83 \citep[see also][]{Korn2012}. 
As an alternative to RGB stars, \citet{gao_galah_2020} have identified pristine Li abundances in warm main-sequence field stars with masses above $1.3\,\Msol$ (i.e., significantly younger than M4) that appear to have neither undergone depletion, nor been enhanced by Galactic chemical evolution. They identified a small number of moderately metal-poor field stars with $-1.0 < {\rm [Fe/H]} < -0.5$ in this group, and found that these exhibit surface abundances $\Abund{Li} = 2.69 \pm 0.06$ that are compatible with BBN predictions. 

We have not accounted for Galactic production of lithium when deriving the initial Li abundance content of the cluster. The empirical trends of Li abundance with metallicity are found to vary in the literature: \citet{Ryan1999} and \citet{Asplund2006} found trends as steep as 0.1\,dex per 1\,dex in $\FeH$, \citet{Melendez2004} and \citet{Shi2007} found no trend at all, \citet{bensby_exploring_2018} found opposite slopes in the thin and thick disks of the Galaxy, and \citet{romano_gaia-eso_2021} identified an additional trend with Galactocentric distance. 
On the theoretical side, \citet{prantzos_production_2012} predicts a (Galactic) production of merely 0.05\,dex at this metallicity, due mainly to $\nu$-nucleosynthesis in core-collapse supernovae rather than spallation by cosmic rays, while \citet{fields_implications_2022-1} predict a cosmic ray production of close to 0.1\,dex. 
Accounting for post-primordial production by applying, e.g., an 0.1\,dex downward correction of the derived stellar lithium abundance would further weaken the agreement with the CMB-calibrated BBN primordial Li value. Such a revision would only be consistent with a more efficient AddMix, as discussed above.
There are also numerous non-standard BBN scenarios that could bridge this gap, e.g., a higher value of the fine-structure constant by only a few ppm at the BBN time \citep{ClMa2020,DeMa2021}.

Another possibility for the low stellar abundances compared to the BBN-predicted primordial Li abundance was suggested by \citet{Piau2006}. They argue that part of the discrepancy of order 0.2--0.3 dex 
is explained by Population III stars that efficiently depleted lithium. This scenario was criticised by \citet{prantzos_early_2006-1} who argued that even a slight depletion of lithium would likely be accompanied by prohibitively large oxygen production in these stars. 
Furthermore, one would expect the amount of mixing through Population III stars to vary depending on the mass of the parent galaxy. Instead, lithium abundances in the Sagittarius globular cluster M54, in the remnant dwarf galaxy $\omega$\,Centauri, and in accreted stars, are similar to those found in Galactic field stars \citep{monaco_lithium_2010,mucciarelli_cosmological_2014,simpson_galah_2021}.

\subsection[The Teff scale]{The $\Teff$\ scale}\label{sec:Tempscales}
The magnitude of the abundance trends is a topic of debate and as shown by the discussion in Sect.~\ref{Sec:teffscale} susceptible to errors in $\Teff$. To check whether the trends are spurious results of potential biases in the temperature scale, we executed the analysis on two other temperature scales. 
We refer to our main temperature scale, derived from the $(V-I)$--$\Teff$ relations of \citet{Ramirez2005}, as $T_{\text{eff, phot}}$, 
the spectroscopic $\Teff$ scale constructed to uphold the ionisation equilibrium between \ion{Fe}{i} and \ion{Fe}{ii} as $T_{\text{eff, ion}}$,
and the spectroscopic $\Teff$ scale constructed to produce a flat abundance trend deduced from \ion{Fe}{i} as $T_{\text{eff, flat}}$. We remind the reader that our reported abundance trend in $\rm [Fe/H]$ is based on lines of \ion{Fe}{ii}, which are less sensitive to changes in $\Teff$.

The spectroscopic and photometric $\Teff$ scales are affected by different types of biases. For example, the photometric $\Teff$ scale obtained via calibration of observed photometric colours on the infrared flux method (IRFM) is largely insensitive to uncertainties in model atmospheres. 
It can, however, be affected by uncertainties in the photometry, non-linearities and discontinuities in the response of $\Teff$\ to photometric colour or chemical composition, lacking or uneven coverage in parameter space, uncertainties in reddening, etc. 
The spectroscopic $\Teff$\ scales are sensitive to modelling shortcomings for model atmospheres and line formation, as well as the quality of the spectra and the completeness of line lists.

\begin{figure}
\begin{center}
\includegraphics[]{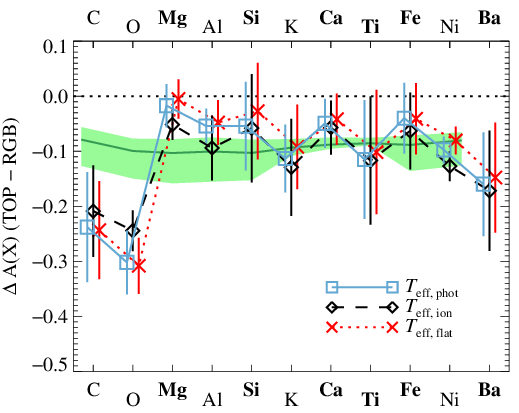}
\caption{Elemental abundance trends on three different $\Teff$\ scales including our primary scale $T_{\text{eff, phot}}$. Abundances were determined from group-averaged coadded spectra, and are shown connected by lines for clarity, with vertical lines indicating the standard deviation. 
Error bars shown are the statistical errors, added in quadrature for the TOP and RGB abundance measurements.
The abundance of iron is based on \ion{Fe}{ii} lines.
The shaded background indicates the range of model predictions for different efficiencies of AddMix, with the T6.2 model indicated by a line.
}\label{Fig:Teff-scales}
\end{center}
\end{figure}

Results of the three abundance analyses are compared in Table~\ref{Tab:Teff-scales}, and visualised in Fig.~\ref{Fig:Teff-scales}, where we also compare to model predictions. 
Compared to the photometric temperature scale, abundance trends on the spectroscopic $\Teff$\ scales $T_{\text{eff, ion}}$ and $T_{\text{eff, flat}}$ are systematically stronger and weaker, respectively, for magnesium, aluminium, potassium, calcium, nickel and barium, while carbon and oxygen show the opposite pattern. The abundances of silicon and of iron and titanium (based on ionised lines) remain essentially unchanged. 

In summary, the trends in carbon, oxygen, aluminium, potassium, calcium, nickel and barium remain formally statistically significant, in that they deviate from no variation between TOP and RGB, by at least 1$\sigma$ on all three $\Teff$\ scales.
This, in line with the reasoning in Sect.~\ref{sec:uncertainties}, verifies that the observed abundance trends are indeed robust to uncertainties in the stellar parameters.
\begin{table*}
\caption{Elemental abundance trends, derived on different $\Teff$ scales including our primary scale $T_{\text{eff, phot}}$, and from stellar evolution models with three different values $\text{T}X$ for AddMix. Differences in abundances are based on the coadded spectra, and are given in the sense TOP--RGB. }
\label{Tab:Teff-scales}
\centering
\begin{tabular}{lrrrrrrr}
\hline
Parameter	 & \multicolumn{1}{c}{$T_{\text{eff, phot}}$} & \multicolumn{1}{c}{$T_{\text{eff, ion}}$} & \multicolumn{1}{c}{$T_{\text{eff, flat}}$} & \multicolumn{1}{c}{T6.0} & \multicolumn{1}{c}{T6.2} & \multicolumn{1}{c}{T6.25} & \multicolumn{1}{c}{T6.3} \\
\hline
$\Delta \Teff$ & $1113\pm198$ & $1063\pm169$ & $1132\pm177$ &  \dots  &  \dots  &  \dots  &  \dots  \\
$\Delta \logg$ & $ 1.28\pm0.46$ & $ 1.27\pm0.45$ & $ 1.29\pm0.46$ &  \dots  &  \dots  &  \dots  &  \dots  \\
$\Delta \vmic$ & $ 0.35\pm0.13$ & $ 0.29\pm0.12$ & $ 0.35\pm0.12$ &  \dots  &  \dots  &  \dots  &  \dots  \\
$\Delta\Abund{Li}^a$         & $ 1.30\pm0.12$ & $ 1.24\pm0.15$ & $ 1.31\pm0.14$ & $+1.21$ & $+1.20$ & $+1.14$ & $+1.03$ \\
$\Delta\Abund{C}$            & $-0.24\pm0.10$ & $-0.21\pm0.08$ & $-0.24\pm0.09$ & $-0.13$ & $-0.08$ & $-0.07$ & $-0.06$ \\
$\Delta\Abund{O}$            & $-0.30\pm0.06$ & $-0.24\pm0.04$ & $-0.31\pm0.05$ & $-0.15$ & $-0.10$ & $-0.09$ & $-0.08$ \\
$\Delta\Abund{Mg}$           & $-0.02\pm0.04$ & $-0.05\pm0.03$ & $-0.01\pm0.04$ & $-0.16$ & $-0.10$ & $-0.09$ & $-0.08$ \\
$\Delta\Abund{Al}$           & $-0.05\pm0.03$ & $-0.09\pm0.06$ & $-0.05\pm0.04$ & $-0.16$ & $-0.10$ & $-0.09$ & $-0.08$ \\
$\Delta\Abund{Si}$           & $-0.05\pm0.08$ & $-0.06\pm0.10$ & $-0.03\pm0.09$ & $-0.15$ & $-0.10$ & $-0.09$ & $-0.08$ \\
$\Delta\Abund{K}$            & $-0.11\pm0.06$ & $-0.13\pm0.09$ & $-0.09\pm0.08$ & $-0.11$ & $-0.09$ & $-0.09$ & $-0.08$ \\
$\Delta\Abund{Ca}$           & $-0.05\pm0.04$ & $-0.06\pm0.05$ & $-0.04\pm0.05$ & $-0.10$ & $-0.09$ & $-0.08$ & $-0.08$ \\
$\Delta\Abund{Ti}$           & $-0.11\pm0.11$ & $-0.12\pm0.12$ & $-0.10\pm0.11$ & $-0.09$ & $-0.09$ & $-0.08$ & $-0.08$ \\
$\Delta\Abund{\ion{Fe}{i}}$  & $-0.02\pm0.03$ & $-0.04\pm0.02$ & $ 0.01\pm0.01$ & $-0.14$ & $-0.09$ & $-0.08$ & $-0.07$ \\
$\Delta\Abund{\ion{Fe}{ii}}$ & $-0.04\pm0.06$ & $-0.06\pm0.07$ & $-0.04\pm0.06$ & $-0.14$ & $-0.09$ & $-0.08$ & $-0.07$ \\
$\Delta\Abund{Ni}$           & $-0.10\pm0.02$ & $-0.13\pm0.03$ & $-0.08\pm0.03$ & $-0.13$ & $-0.08$ & $-0.07$ & $-0.07$ \\
$\Delta\Abund{Ba}$           & $-0.16\pm0.09$ & $-0.17\pm0.11$ & $-0.15\pm0.10$ &  \dots  &  \dots  &  \dots  &  \dots  \\
\hline
\end{tabular}\\
\textit{(a)}: Average of individual star abundances rather than group averages, with $T_{\text{eff, phot}} > 5900$\,K (TOP) and $4800\,\text{K} < T_{\text{eff, phot}} < 5200\,\text{K}$ (RGB).
\end{table*}

\begin{figure*}
\begin{center}
\includegraphics[]{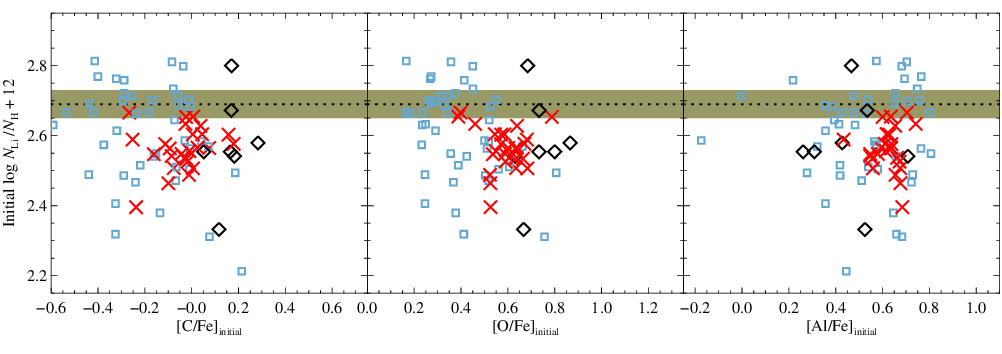}
\caption{Abundances of lithium compared to the abundance ratios over iron of carbon (left), oxygen (middle) and aluminium (right). All abundances have been corrected for the evolutionary effects of atomic diffusion and dredge-up. The predicted primordial lithium abundance based on CMB-calibrated BBN calculations (see text) is indicated by the dotted line and shaded region.
Symbols and colours are the same as in Fig.~\ref{Fig:CNO}, with small blue squares representing dwarfs, and red crosses and black diamonds representing N-poor and N-rich giants, respectively.
}\label{Fig:lithium_anticor}
\end{center}
\end{figure*}

\subsection{The formation of M4}\label{sec:formation}
M4 is characterised by a bimodal N distribution and bimodal Na-O anti-correlation \citep{Villanova2011}, a C-O correlation and N-O anti-correlation \citep{Ivans1999}, and the lack of an Mg-Al anti-correlation \citep[][]{Villanova2011}. 
The simplest explanation for the observed abundance patterns is a self-pollution scenario with two or more distinct star formation epochs, where the first generation formed out of primordial (O-rich, N- and Na-poor) material and polluted the medium before the later generations of stars formed. 
The origin of the pollution is debated. The most common candidates include intermediate-mass AGB stars \citep{ventura_predictions_2001}, fast-rotating massive stars \citep{Decressin2007a,Decressin2007b}, massive binary stars \citep{deMink2009}, novae \citep{maccarone_novae_2012} and super-massive stars \citep{gieles_concurrent_2018}. Beyond the scenarios with multiple star-formation epochs, we note also the suggestion of late time accretion amongst coeval low-mass stars \citep{bastian_early_2013}.

The fact that M4 does not show evidence of a Mg-Al anti-correlation indicates that the Mg-Al burning cycle was not active in the polluting stars, implying that they did not reach core temperatures of $50 \times 10^6$\,K that are required for the Mg-Al burning cycle. 
We use models from \citet{Decressin2007a} to derive an upper limit to the mass of fast-rotating massive stars (FRMS) in this scenario. Based on our range of $\rm [Mg/Al]$ ratios, we determine an upper limit to the mass of 20--40\,$\Msol$. This is consistent with the upper-mass limit given by \citet{Villanova2011} based on the range in their $\rm [O/Na]$ ratios. Disregarding our very lowest abundances $\XY N {Fe}$, our observed range in $\rm [C/N]$ is in line with predictions for the 40\,$\Msol$ model. This, together with the observed N-Al and Al-Si correlations, seems to suggest that the Mg-Al cycle was in fact active, especially since the Al-Si correlation is a direct result of leakage from the Mg-Al cycle on $^{28}$Si which requires a temperature of at least $65\times10^6$\,K \citep{Carretta2009a}. Evidence for an active Mg-Al cycle was also presented by \citet{Marino2011}.
This suggests that the pollution scenario involves pollution by FRMS with masses of roughly 40\,$\Msol$. However, the high [C/N] values detected in the N-poor stars seem to suggest that there is some other pollution mechanism at work as well.

\citet{Villanova2011} ruled out a pollution scenario in which the second generation of stars was born from material polluted by AGB stars. They found that the barium abundances they derived for a group of RGB stars in M4 do not show a bimodal behaviour, while the yttrium abundances do. They argued that, since their observations are not in line with AGB yields calculated by \citet{Karakas2010} which indicate a similar behaviour for the s-process elements if the pollution is driven by massive AGB stars, the AGB scenario is not plausible. 
\citet{meszaros_homogeneous_2020} argue the s-process enhancement must be unrelated to the light element abundance variations, i.e. the pollution must have been introduced after the clusters had already formed \citep[see also][]{masseron_homogeneous_2019}. 

We note that AGB stars may be able to produce Li through the \citet{Cameron1971} mechanism. We have used the T6.2 stellar evolution models to remove the influence of AD on surface abundances to derive the initial abundances of Li, C, O and Al for our full sample of dwarfs and giants, and present these in Fig.~\ref{Fig:lithium_anticor}. We do not uncover any clear (anti-)correlations with Li for any of the three elements, nor any systematic variation with N in the giant stars. It was argued by \citet{dorazi_lithium_2010} that the lack of correlations could be a sign of Li production. While AGB stars would be the most likely candidate for this, it should be noted that the Cameron-Fowler mechanism is susceptible to assumptions on mass-loss rates and quite some fine-tuning would be required to achieve uniform Li abundances in our sample. 

In light of the findings here and in the literature we suggest a scenario in which the pollution is caused by both FRMS and AGB stars. We can envision a scenario where the massive stars $(\sim40\,{\rm M}_{\odot})$ of the first generation are FRMS which are responsible for the initial pollution. As the evolution proceeds, intermediate-mass stars $(\sim 10\,{\rm M}_{\odot})$ enter the AGB phase and are responsible for another injection of processed material. 
To get to the bottom of this, we suggest a new study in which one combines information on s-process abundances in a homogeneous analysis with (anti-)correlation information on the light elements.

\section{Conclusions} \label{sec:summary}
Our chemical abundance analysis indicates the existence of weak abundance trends along the subgiant branch in magnesium, silicon, calcium, titanium and iron. We find that these trends are robust to modelling uncertainties, as well as uncertainties in the $\Teff$\ scale. The observed trend in iron would, e.g., require changes of several hundred kelvin to flatten completely. Additionally, the trends are found to be in very good agreement with predictions from stellar structure models including atomic diffusion (AD) moderated by efficient additional mixing (AddMix). 
We also find statistically significant trends in carbon, oxygen, aluminium, potassium, nickel and barium, which are robust to uncertainties in the $\Teff$\ scale. We caution that some of these elements, such as K, may be significantly distorted by differential NLTE effects. 

In the current formulation of the AddMix mechanism, its efficiency needs to be at least T6.2 in order to reproduce the observed trends. This is in agreement with results from NGC\,6752 \citep[$\FeH = -1.6$,][]{Gruyters2013,Gruyters2014} and M30 \citep[$\FeH = -2.3$][]{gavel_atomic_2021}, but contrary to NGC\,6397 \citep[$\FeH = -2.1$,][]{Korn2007,Lind2008,Nordlander2012} where a weaker AddMix efficiency of T6.0 is required to match observations.
Several open clusters with near-solar metallicity have been analysed in the literature. M67 (4\,Gyr, $\FeH = 0.0$) has been analysed by several authors: 
\citet{onehag_abundances_2014} found weak but systematic abundance differences between TOP and MS (using the solar twin M67-1194) at the level of just 0.03\,dex, in excellent agreement with stellar evolution models without turbulent mixing;
\citet{gao_galah_2018} and \citet{liu_chemical_2019} found abundance differences of roughly 0.1\,dex between TOP and SGB, which were in good agreement with the atomic diffusion models from \citet{dotter_influence_2017};
\citet{bertelli_motta_gaia-eso_2018-1} and \citet{souto_chemical_2018,souto_chemical_2019} found abundance differences of typically 0.1--0.2\,dex between MS and RGB stars, in broad agreement with atomic diffusion both with and without AddMix. 
\citet{semenova_gaia-eso_2020} analysed NGC~2420 (2.6\,Gyr, $\FeH = -0.05$), finding that TOP stars were depleted by as much as 0.2\,dex relative to lower-MS and RGB stars, in agreement with predictions with weak AddMix (T5.8). 
In binary field stars with broadly solar metallicity, \citet{liu_detailed_2021} found small but significant star-to-star abundance variations of a few 0.01\,dex, in good agreement with \citet{dotter_influence_2017}. 
In contrast to these solar-metallicity clusters that match predictions with weak or no AddMix, the solar \Abund{Li} value is well predicted by a model with strong additional mixing \citep[T6.2: ][]{Richard2002}. It is thus not clear how the trend in AddMix continues from $\FeH = -1.1$ toward solar metallicity, and if there are other parameters to consider  (e.g.\ stellar rotation).

After correcting our measured lithium abundances for the predicted effects of AD and dredge-up, we determine an average initial lithium abundance of $A(\text{Li})_\text{init} = 2.63 \pm 0.10$. 
We note that results from the four globular clusters indicate consistent diffusion-corrected initial lithium abundances, in the very narrow range $A(\text{Li})_\text{init} = 2.48$ \citep{gruyters_atomic_2016} to 2.63 (this work), fully compatible with each other within the associated errors.

In order to constrain the properties of first-generation polluters in the cluster, we have compared abundances of elements that form under different conditions. 
The observed ranges of abundance ratios $\XY{Mg}{Al}$ and  $\XY{C}{N}$ are consistent with an upper mass limit for the polluting stars of roughly 40\,$\Msol$ \citep{Decressin2007a}, in broad agreement with what \citet{Villanova2011} deduced from the same theoretical models using the corresponding range in $\XY O{Na}$. 
We cannot, however, reconcile our non-detection of a Mg-Al anti-correlation with the detected N-Al and Al-Si correlations which indicate leakage from an active Mg-Al cycle. 
We thus ask stellar modellers to further investigate possible evolutionary scenarios which could generate these abundance patterns.

\section*{Acknowledgements}
The authors thank the anonymous referee for their helpful comments and suggestions, which have improved the manuscript.
We also thank Yeisson Osorio, Lyudmila Mashonkina, Tatyana Sitnova and Sofia Alexeeva for performing NLTE calculations specifically for this work, and for their patience over the years with seeing this work finally be published.
We thank Yassan Momany for providing the photometric catalogue of M4, and Vidas Dobrovolskas for providing us with tabulations of 3D abundance corrections. 
Parts of this research were supported by the Australian Research Council Centre of Excellence for All Sky Astrophysics in 3 Dimensions (ASTRO 3D), through project number CE170100013. TN and AK acknowledge support by the Swedish National Space Board. 
PG and AK thank the European Science Foundation for support in the framework of EuroGENESIS. 
O.R. acknowledges MESO@LR and Calcul Qu\'ebec for providing the computational resources required for the stellar evolutionary computations. O.R. also acknowledges the financial support of Programme National de Physique Stellaire (PNPS) of CNRS/INSU.
This work was supported by computational resources provided by the Australian Government through the National Computational Infrastructure (NCI) under the National Computational Merit Allocation Scheme (project y89).
This work is based on observations made with ESO Telescopes at the La Silla Paranal Observatory under programme ID 081.D-0356. 

\section*{Data Availability}
The spectroscopic data used in this work are publicly available via the ESO archive facility, under programme ID 081.D-0356. 
The processed spectroscopic data used in this work are supplied on reasonable request to the corresponding author. 

\bibliographystyle{mnras}
\bibliography{allreferences_M4,zotero-library-abbr}

\appendix
\section{Stellar parameters and abundances for the group averages}
Average stellar parameters and the abundances derived from co-added group spectra are provided in Table~\ref{Tab:coadd-results}. This includes the average difference in abundance between the TOP and RGB spectra. We note that for Li, the RGB abundance is based only on RGB2 and RGB3, as stars in RGB1 are located above the RGB bump and thus affected by an additional mixing episode \citep[see][]{Lind2009}.

\begin{landscape}
\begin{table}
\caption{Derived elemental abundances for the coadded group-averaged spectra. Abundance uncertainties are based on the statistical error as calculated by SME. The final column gives the abundance difference between TOP and RGB. The TOP and RGB abundances are the averages of results for the three respective coadded group-averaged spectra.
The uncertainty on the trends is based on the standard deviation of the two averages. }
\label{Tab:coadd-results}
\centering
\begin{tabular}{lccc cccc ccc }
\hline
	 & RGB1 	 & RGB2 	 & RGB3 	    & SGB1     & SGB2     & SGB3 		& TOP1 		& TOP2 		 & TOP3 & $\Delta$(TOP--RGB) \\
\hline 
$\Teff$\ (K) & $  4777$ & $  5000$ & $  5085$ & $  5388$ & $  5636$ & $  5728$ & $  5973$ & $  5995$ & $  6084$ & $  1063$ \\
$\log g$ (cgs) & $ 2.32$ & $ 2.89$ & $ 3.21$ & $ 3.75$ & $ 3.87$ & $ 3.93$ & $ 4.01$ & $ 4.06$ & $ 4.14$ & $ 1.26$ \\ 
$\Abund{Li}$     & $ 0.71\pm 0.08$ &  $ 0.99\pm 0.08$ &  $ 1.10\pm 0.06$ &  $ 1.41\pm 0.08$ &  $ 1.97\pm 0.04$ &  $ 2.09\pm 0.02$ &  $ 2.40\pm 0.03$ &  $ 2.32\pm 0.03$ &  $ 2.32\pm 0.03$ & \phantom{$-$}$ 1.43\pm 0.21$  \\
$\Abund{C}$      & $ 7.18\pm 0.07$ &  $ 7.25\pm 0.06$ &  $ 7.18\pm 0.05$ &  $ 7.25\pm 0.06$ &  $ 7.16\pm 0.05$ &  $ 7.09\pm 0.03$ &  $ 7.05\pm 0.05$ &  $ 6.91\pm 0.04$ &  $ 7.02\pm 0.03$ & $-0.24\pm 0.10$  \\ 
$\Abund{N}$      & $ 7.67\pm 0.07$ &  $ 7.59\pm 0.06$ &  $ 7.58\pm 0.06$ &  $ 7.79\pm 0.08$ &  \dots &  \dots &  \dots &  \dots &  \dots & \dots \\
$\Abund{O}$      & $ 8.13\pm 0.05$ &  $ 8.11\pm 0.04$ &  $ 8.16\pm 0.03$ &  $ 8.16\pm 0.04$ &  $ 7.93\pm 0.05$ &  $ 8.04\pm 0.02$ &  $ 7.92\pm 0.04$ &  $ 7.88\pm 0.02$ &  $ 7.87\pm 0.02$ & $-0.27\pm 0.04$  \\
$\Abund{Mg}$     & $ 6.78\pm 0.02$ &  $ 6.79\pm 0.02$ &  $ 6.78\pm 0.02$ &  $ 6.72\pm 0.04$ &  $ 6.64\pm 0.05$ &  $ 6.71\pm 0.03$ &  $ 6.70\pm 0.06$ &  $ 6.73\pm 0.04$ &  $ 6.72\pm 0.04$ & $-0.05\pm 0.01$  \\ 
$\Abund{Al}$     & $ 5.77\pm 0.01$ &  $ 5.76\pm 0.01$ &  $ 5.69\pm 0.02$ &  $ 5.68\pm 0.04$ &  $ 5.59\pm 0.05$ &  $ 5.61\pm 0.04$ &  $ 5.57\pm 0.08$ &  $ 5.54\pm 0.06$ &  $ 5.54\pm 0.05$ & $-0.19\pm 0.05$  \\ 
$\Abund{Si}$     & $ 6.75\pm 0.02$ &  $ 6.73\pm 0.01$ &  $ 6.71\pm 0.01$ &  $ 6.70\pm 0.03$ &  $ 6.77\pm 0.03$ &  $ 6.72\pm 0.02$ &  $ 6.72\pm 0.03$ &  $ 6.60\pm 0.03$ &  $ 6.68\pm 0.02$ & $-0.06\pm 0.06$  \\ 
$\Abund{K}$      & $ 4.62\pm 0.02$ &  $ 4.63\pm 0.02$ &  $ 4.57\pm 0.02$ &  $ 4.56\pm 0.04$ &  $ 4.61\pm 0.04$ &  $ 4.62\pm 0.03$ &  $ 4.57\pm 0.05$ &  $ 4.50\pm 0.03$ &  $ 4.44\pm 0.03$ & $-0.10\pm 0.08$  \\
$\Abund{Ca}$     & $ 5.45\pm 0.01$ &  $ 5.46\pm 0.01$ &  $ 5.45\pm 0.01$ &  $ 5.40\pm 0.02$ &  $ 5.46\pm 0.01$ &  $ 5.43\pm 0.01$ &  $ 5.42\pm 0.02$ &  $ 5.38\pm 0.01$ &  $ 5.35\pm 0.01$ & $-0.07\pm 0.04$  \\ 
$\Abund{Ti}$     & $ 4.09\pm 0.02$ &  $ 4.12\pm 0.02$ &  $ 4.17\pm 0.02$ &  $ 3.97\pm 0.07$ &  $ 4.08\pm 0.06$ &  $ 4.05\pm 0.05$ &  $ 4.27\pm 0.10$ &  $ 4.29\pm 0.07$ &  $ 4.43\pm 0.06$ & $-0.11\pm 0.11$  \\ 
$\Abund{\ion{Fe}{ii}}$  & $ 6.24\pm 0.01$ &  $ 6.26\pm 0.00$ &  $ 6.27\pm 0.01$ &  $ 6.23\pm 0.01$ &  $ 6.31\pm 0.01$ &  $ 6.25\pm 0.01$ &  $ 6.25\pm 0.02$ &  $ 6.20\pm 0.01$ &  $ 6.20\pm 0.01$ & $-0.07\pm 0.07$  \\ 
$\Abund{Ni}$      & $ 5.07\pm 0.01$ &  $ 5.07\pm 0.01$ &  $ 5.06\pm 0.01$ &  $ 4.99\pm 0.02$ &  $ 5.06\pm 0.02$ &  $ 5.01\pm 0.01$ &  $ 5.04\pm 0.03$ &  $ 4.97\pm 0.02$ &  $ 4.96\pm 0.02$ & $-0.08\pm 0.04$  \\ 
$\Abund{Ba}$     & $ 1.43\pm 0.02$ &  $ 1.34\pm 0.02$ &  $ 1.31\pm 0.02$ &  $ 1.30\pm 0.04$ &  $ 1.25\pm 0.04$ &  $ 1.25\pm 0.03$ &  $ 1.27\pm 0.04$ &  $ 1.09\pm 0.03$ &  $ 1.11\pm 0.03$ & $-0.20\pm 0.12$  \\
\hline
\end{tabular}
\end{table}
\end{landscape}

\bsp	
\label{lastpage}
\end{document}